\documentclass[12pt]{article}
\usepackage{epsfig}
\usepackage{psfrag}
\usepackage{enumitem}
\usepackage{latexsym}
\usepackage{indentfirst}
\usepackage{fancyhdr}
\usepackage{amssymb}
\usepackage{amsmath}
\usepackage{amsfonts}
\usepackage{pifont}
\usepackage{cite}
\usepackage{bbold}
\usepackage{ulem}
\usepackage{color}
\usepackage[footnotesize]{caption2}
\usepackage{graphicx}
\usepackage[center,footnotesize,hang]{subfigure}
\usepackage{url}
\usepackage[dvipsnames]{xcolor}
\usepackage{array}
\usepackage{slashed}

\begin{document}
\title{ \hfill\mbox{\small SISSA 01/2015/FISI}\\[-3mm]
\hfill ~\\[0mm]
       \textbf{A Bottom-Up Approach to Lepton Flavor and CP Symmetries}        }
\date{}
\author{\\[1mm]Lisa L.~Everett$^{1,2\,}$\footnote{E-mail: {\tt
leverett@wisc.edu}}~,~
Todd~Garon$^{1\,}$\footnote{E-mail: {\tt tgaron@wisc.edu}}~,
Alexander J.~Stuart$^{3,4\,}$\footnote{E-mail: {\tt astuart@sissa.it}}\\
\\[1mm]
  \it{\small $^1$Department of Physics, University of Wisconsin,}\\
  \it{\small Madison, WI 53706, USA}\\[4mm]
\it{\small $^2$ Enrico Fermi Institute, University of Chicago,}\\
  \it{\small Chicago, IL 60637, USA}\\[4mm]
  \it{\small $^3$School of Physics and Astronomy, University of
  Southampton,}\\
  \it{\small Southampton, SO17 1BJ, United Kingdom}\\[4mm]
 \it{\small $^4$SISSA/INFN, Via Bonomea 265, 34136 Trieste, Italy}\\
 }

\maketitle
\thispagestyle{empty}
\begin{abstract}
\noindent
We perform a model-independent analysis of the possible residual Klein and generalized  CP symmetries associated with arbitrary lepton mixing angles in the case that there are three light Majorana  neutrino species. This approach emphasizes the unique role of the Majorana phases and provides a useful framework in which to discuss the origin of the Dirac CP phase in scenarios with spontaneously broken flavor and generalized CP symmetries.  The method is shown to reproduce known examples in the literature based on tribimaximal and bitrimaximal mixing patterns, and is used to investigate these issues for the case of a particular (GR1) golden ratio mixing pattern.

\end{abstract}

\newpage
%\setcounter{page}{1}

%%%%%%%%%%%%%%%%%%%%%%%%%%%%%%%%

\section{Introduction\label{sec:intro}}

%%%%%%%%%%%%%%%%%%%%%%%%%%%%%%%%

The recent measurement of the reactor mixing angle by the Daya Bay\cite{dayabay}, RENO\cite{reno}, and Double Chooz\cite{doublechooz} collaborations has heralded the beginning of the age of precision lepton measurements and has opened the door for direct observation of CP violation in the lepton sector.  With the assumption of three light neutrino species, the pattern of the lepton mixing angles of the Maki-Nakagawa-Sakata-Pontecorvo (MNSP) lepton mixing matrix, $U_{\rm MNSP}$ \cite{pdg} is now on firm experimental ground, with two large angles associated with atmospheric and solar neutrino oscillations, and the reactor mixing angle, which is of the order of the Cabibbo angle of the quark sector, modulo $O(1)$ factors.  While there are no direct experimental limits on the CP-violating phases of $U_{\rm MNSP}$, global fits to lepton mixing parameters \cite{global} may already be providing compelling hints for the existence of ``Dirac"-type CP violation in the lepton mixing, begging theorists to be prepared for its possible future measurement. 

Of the possible theoretical approaches to explaining the origin of lepton family mixing, perhaps the most provocative in the case that neutrinos are Majorana particles is the assumption of a high scale discrete flavor symmetry group that is spontaneously broken to a residual Klein symmetry at low energies.  This residual Klein symmetry completely fixes the elements of $U_{\rm MNSP}$ in the diagonal charged lepton basis (up to charged lepton rephasing), though it fails to provide concrete predictions for the Majorana phases, $\alpha_{21}$ and $\alpha_{31}$ \cite{pdg}.  However, recent work\cite{lindner} concerning the \textit{consistent} implementation of a generalized CP symmetry alongside such a flavor symmetry has changed this situation.  As such many models and analyses of CP and flavor symmetries have been studied, e.g.~$A_4$\cite{A4CP}/$\Delta(3n^2)$\cite{Ding:2014hva,Hagedorn:2014wha}, $S_4$\cite{Feruglio:2012cw,S4CP,Feruglio:2013hia,Li:2013jya,bimaxS4}/$\Delta(6n^2)$\cite{Hagedorn:2014wha,NandK,Ding:2014ssa,Ding:2014ora}, $\Sigma(n\phi)$\cite{sigmagroups}, and $T'$\cite{tprime}.\footnote{CP has also been studied ``model-independently'' with a single preserved $Z_2$ residual neutrino flavor symmetry\cite{Luhn:2013lkn,Z2CP1,Z2CP2,Z2CP3,Z2CP4}.}   In this context, the goal of the present work is not to construct a specific top-down model of this type, but instead to understand how the measured lepton mixing parameters shape the residual CP and flavor symmetries from a bottom-up perspective, which can provide a useful guideline for theoretical model-building within this general framework.\footnote{Refs.~\cite{scans0,scans1} have used similar methods as described in this work to find flavor symmetry groups containing viable flavor subgroups.}

This paper is structured as follows.  In Section \ref{sec:klein}, after reviewing the existence of a residual Klein symmetry for Majorana neutrinos, we determine its group elements and the general Majorana neutrino mass matrix as functions of the observed leptonic mixing parameters.  In Section \ref{sec:genCP}, the residual CP symmetries consistent with such a remnant Klein symmetry are found and expressed in terms of mixing angles and CP-violating phases.  We also discuss the possible origins of nontrivial CP-violating phases for scenarios in which generalized CP and flavor symmetries are nontrivially connected.  In Section \ref{sec:applications}, we demonstrate how our model-independent results reproduce key results in previous literature, and  also elucidate a new example based on a particular golden ratio mixing pattern\cite{GRmixing0,GRmixing1} that can be obtained e.g.~in $A_5$ models\cite{A5GRmodels}.  Finally in Section \ref{sec:conclusion}, we summarize and present our conclusions.

%%%%%%%%%%%%%%%%%%%%%%%%%%%%%%%%

\section{Klein Symmetry and the Neutrino Sector\label{sec:klein}}

%%%%%%%%%%%%%%%%%%%%%%%%%%%%%%%%

We begin with an overview of the Klein symmetry in lepton mixing, which relies on the assumption that the neutrinos are Majorana fermions (for a review, see Ref.~\cite{kingluhnreview}).  If neutrinos are Majorana, the neutrino mass matrix $M_\nu$ is complex symmetric, {\it i.e.}, $M_{\nu}=M_{\nu}^T$, and hence can be diagonalized by a unitary matrix, $U_{\nu}$, as follows:
\begin{equation}
\label{eq:DiagMnu}
U^T_{\nu}M_{\nu}U_{\nu}=M^{\text{Diag}}_{\nu}=\text{Diag}(m_1,m_2,m_3)=\text{Diag}(|m_1| e^{-i\alpha_1},|m_2| e^{-i\alpha_2},|m_3|e^{-i\alpha_3}),
\end{equation}
in which the $\alpha_{1,2,3}$ are Majorana phases \cite{majorana1,majorana2}.  Here we emphasize that for reasons which will become apparent shortly, we choose to leave the neutrino mass eigenvalues as complex parameters, and hence our expression for $U_{\nu}$ is related to the standard version in which the neutrino masses are real and positive by the replacement of $U_\nu \rightarrow U_\nu P_{\rm Maj}$, in which $P_{\rm Maj}= \text{Diag}(e^{i \alpha_1/2}, e^{i\alpha_2/2}, e^{i \alpha_3/2})$.

Notice the lack of complex conjugation in Eq.~(\ref{eq:DiagMnu}).  This is in contrast to the diagonalization of the charged lepton mass matrix $m_e$, or more precisely of the Hermitian combination $M_e=m_em_e^{\dagger}$, which takes the form 
\begin{equation}
\label{eq:DiagMe}
U_e^{\dagger}M_e U_e=M^{\text{Diag}}_e=\text{Diag}(|m_e|^2,|m_{\mu}|^2,|m_{\tau}|^2).
\end{equation}
Eqs.~(\ref{eq:DiagMnu})-(\ref{eq:DiagMe}) have suppressed unphysical phase redundancies that can be restored with the introduction of phase matrices
\begin{equation}\label{eq:redphase} 
Q_e=\text{Diag}(e^{i\beta_1},e^{i\beta_2},e^{i\beta_3}),~Q_{\nu}=\text{Diag}(\pm 1, \pm 1, \pm 1),
\end{equation}
in which the $\beta_{1,2,3}$ are arbitrary phases associated with the freedom to rephase the charged lepton fields, and the entries of $Q_{\nu}$ are constrained by the lack of complex conjugation in Eq.~(\ref{eq:DiagMnu}) to be only $\pm 1$.  Hence, Eq. (\ref{eq:DiagMnu}) can be re-expressed as 
\begin{equation}\label{eq:DiagMnu2}
M^{\text{Diag}}_{\nu}=Q_{\nu}^TU^T_{\nu}M_{\nu}U_{\nu}Q_{\nu},
\end{equation}
while Eq. (\ref{eq:DiagMe}) becomes
\begin{equation}\label{eq:DiagMe2}
M^{\text{Diag}}_e=Q_e^{\dagger}U_e^{\dagger}M_e U_eQ_e.
\end{equation}
Using Eqs.~(\ref{eq:DiagMnu2})-(\ref{eq:DiagMe2}), the lepton mixing matrix $U_{\rm MNSP}$ then takes the form
\begin{equation}\label{eq:MNSPphases}
U_{\rm MNSP}=Q_e^{\dagger}U_e^{\dagger}U_{\nu} Q_{\nu}.
\end{equation}
From the above, it is tempting to identify the nonzero entries of $Q_{\nu}$ as contributions to the Majorana phases, but we prefer to keep $Q_{\nu}$ separate from $\alpha_{1,2,3}$  so that we can later identify the two phase differences $(\alpha_{2,3}-\alpha_1)/2$ with the Majorana phases of the PDG parametrization of $U_{\rm MNSP}$ \cite{pdg} (recall these phases are encoded here in the neutrino mass eigenvalues).   A further simplification of Eq. (\ref{eq:MNSPphases}) clearly occurs when we work in a basis in which the charged leptons are diagonal. In this case, $U_{\text{MNSP}}=U_{\nu} Q_{\nu}$, up to charged lepton rephasing.  For the time being, we will work for simplicity in the diagonal charged lepton basis, and comment later on the effects of charged lepton corrections, which can occur at a subleading level in many scenarios in the literature. 

With these assumptions,  it is insightful to analyze the residual symmetries of the neutrino sector to understand the possible symmetries involved in lepton mixing.   We begin with the diagonalized Majorana neutrino mass matrix in Eq. (\ref{eq:DiagMnu}), $M_{\nu}^{\text{Diag}}$, which obeys the following relation (as seen from Eq.~(\ref{eq:DiagMnu2})):
\begin{equation}\label{1b2}
M^{\text{Diag}}_{\nu}=Q_{\nu}^TM_{\nu}^{\text{Diag}}Q_{\nu}.
\end{equation}
The eight possible $Q_{\nu}$ thus represent residual symmetry transformations in the basis in which the neutrino mass matrix is diagonal.
Further requiring that these transformations have $\text{Det}(Q_{\nu})=+1$ reduces this set to only four remaining transformations (the set with  $\text{Det}(Q_{\nu})=-1$ is clearly physically redundant): the identity $G_0^{\text{Diag}}=1_{3\times 3}$, and
 \begin{eqnarray}
\label{Gs}
G_1^{\text{Diag}}=\left(
\begin{array}{ccc}
1&0&0\\
0&-1&0\\
0&0&-1
\end{array}\right),\;
G_2^{\text{Diag}}=\left(
\begin{array}{ccc}
-1&0&0\\
0&1&0\\
0&0&-1
\end{array}\right),\;
G_3^{\text{Diag}}=\left(
\begin{array}{ccc}
-1&0&0\\
0&-1&0\\
0&0&1
\end{array}\right).
\end{eqnarray}  
These transformations satisfy
\begin{equation}
\label{klein}
\begin{aligned}
&(G^{\text{Diag}}_i)^2=1 \text{, for $i$ = 0, 1, 2, 3,}\\
&G^{\text{Diag}}_0 G^{\text{Diag}}_i=G^{\text{Diag}}_i\text{, for $i$ =1, 2, 3,}\\
&G^{\text{Diag}}_i G^{\text{Diag}}_j=G^{\text{Diag}}_k \text{, for $i\neq j\neq k\neq 0$}. 
\end{aligned}
\end{equation}
Hence, these matrices furnish a representation of the Klein symmetry group, $K_4\cong Z_2\times Z_2$.  Using this information and the diagonalization condition of Eq.~(\ref{eq:DiagMnu}), it is straightforward to see that the (undiagonalized) neutrino mass matrix $M_{\nu}$ also possesses a Klein symmetry:
\begin{equation}
\label{eq:groupinv}
M_{\nu}=G_i^TM_{\nu}G_i,
\end{equation}
in which the $G_i$ is related to $G_i^{\text{Diag}}$ by
\begin{equation}
\label{k4change}
G_i=U_{\nu}G_i^\text{Diag} U_{\nu}^{\dagger}.
\end{equation}
 Therefore, the $G_i$ form another representation/basis of $K_4$ that is related to the diagonal representation/basis of Eq.~(\ref{Gs}) through Eq.~(\ref{k4change}).\footnote{$G_1$, $G_2$, $G_3$ are also known in the literature as $SU$, $S$, and $U$ respectively.}  Due to the form of the transformation defined in Eq.~(\ref{k4change}), each of the $G_{i=1,2,3}$ has a nondegenerate +1 eigenvalue and two degenerate -1 eigenvalues with the eigenvectors associated with the nondenerate +1 eigenvalues forming columns of $U_{\nu}$.\footnote{We also note that if one applies Eq.~(\ref{k4change}), to the unphysical phase matrices $Q_{\nu}$ with -1 determinant, the resulting matrices are $-G_0, -G_1,-G_2, -G_3$, justifying their earlier dismissal.}  
Therefore, Eq. (\ref{k4change}) provides the mapping of the diagonal $Z_2\times Z_2$ Klein elements to the more ``useful'' basis associated with the leptonic mixing angles of $U_\nu$. This change of basis will allow the use of low energy mixing parameters as inputs to reveal the residual Klein generators, neutrino mass matrix, and allowed generalized CP symmetries.

We begin by discussing the Klein generators.  As previously mentioned, the diagonal elements of the residual low energy Klein symmetry may be transformed into a new set of Klein elements which are functions of the angles in the lepton mixing matrix, $U_\nu$, by virtue of Eq.~(\ref{k4change}). To this end, we will first parametrize $U_\nu$ as follows:
\begin{equation}
\label{eq:symunu}
U_\nu=PR_{x}(\theta_{23},\delta_x)R_{y}(\theta_{13},\delta_y)R_{z}(\theta_{12},\delta_z),
\end{equation}
 in which
\begin{eqnarray}
\label{euler}
\begin{array}{cc}
R_{x}(\theta_{23},\delta_x)=\left(
\begin{array}{ccc}
 1 & 0 & 0 \\
 0 & c_{23} & s_{23}e^{-i \delta_x} \\
 0 & -s_{23}e^{i \delta_x} & c_{23}
\end{array}
\right),&
R_{y}(\theta_{13},\delta_y)=\left(
\begin{array}{ccc}
 c_{13} & 0 & s_{13}e^{-i \delta_y} \\
 0 & 1 & 0 \\
 -s_{13}e^{i \delta_y} & 0 & c_{13}
\end{array}
\right), \\&\\
R_{z}(\theta_{12},\delta_z)=\left(
\begin{array}{ccc}
 c_{12} & s_{12}e^{-i \delta_z} & 0 \\
 -s_{12}e^{i \delta_z} & c_{12} & 0 \\
 0 & 0 & 1
\end{array}
\right),&
P=\left(
\begin{array}{ccc}
1 & 0 & 0 \\
0 & 1 & 0 \\
0 & 0 & -1
\end{array}
\right).
\end{array}
\end{eqnarray}
(The advantage of introducing the matrix $P$ will be discussed shortly.)  Upon expanding Eq.~(\ref{eq:symunu}), $U_\nu$ takes the form
\begin{eqnarray}
\label{MNSP}
U_\nu=
\left(
\begin{matrix}
 c_{12} c_{13} & c_{13} s_{12}e^{-i \delta_z} & s_{13}e^{-i \delta_y} \\
 -c_{23} s_{12}e^{-i \delta_z}-c_{12} s_{13} s_{23}e^{-i (\delta_x-\delta_y)} & c_{12} c_{23}-s_{12} s_{13} s_{23}e^{-i (\delta_x-\delta_y+\delta_z)} & c_{13} s_{23}
e^{-i \delta_x} \\
c_{12} c_{23} s_{13}e^{i \delta_y}-s_{12} s_{23}e^{i(\delta_x+\delta_z)} & c_{23} s_{12} s_{13}e^{i( \delta_y-\delta_z)}+c_{12} s_{23}e^{i \delta_x} & -c_{13} c_{23} \label{Unu}
\end{matrix}
\right),
\end{eqnarray}
in which $s_{ij}=\sin(\theta_{ij})$ and $c_{ij}=\cos(\theta_{ij})$.
In this symmetric parametrization, each mixing angle comes with an associated phase, cf.~Eq.~(\ref{euler}); we choose this as the starting point because it clearly captures the phase degrees of freedom in $U_{\nu}$. However, the ensuing discussion can be simplified with a judicious choice of rephasings, as follows.  First, let us define 
\begin{align*}
P'=\text{Diag}\left(1, \exp\left(-i\delta_z\right), \exp\left(-i(\delta_z+\delta_x)\right)\right).
\end{align*}
Then, it is easy to see that 
\begin{align}
P' U_{\nu}(\theta_{23},\theta_{13},\theta_{12},\delta) P'^* \label{NewParam}=\left(
\begin{array}{ccc}
 c_{12} c_{13} & c_{13} s_{12} &s_{13} e^{-i \delta }  \\
 -c_{23} s_{12}-c_{12}  s_{13} s_{23}e^{i \delta } & c_{12} c_{23}-s_{12} s_{13} s_{23}e^{i \delta }  & c_{13} s_{23} \\
 -s_{12} s_{23}+c_{12} c_{23}  s_{13}e^{i \delta } & c_{12} s_{23}+c_{23}  s_{12} s_{13}e^{i \delta } & -c_{13} c_{23} \\
\end{array}
\right),
\end{align}
in which we have defined the quantity $\delta=\delta_y-\delta_x-\delta_z$.  Notice that left multiplication by $P$ and $P'$ in Eqs.~\eqref{eq:symunu} and \eqref{NewParam} is possible by charged lepton phases redefinition.  Additionally, multiplication by $P'^*$ on the right is equivalent to redefining the Majorana phases on  the (complex) neutrino masses.

Using the above parametrization of $U_\nu$, the definitions for the diagonal generators $G_i^\text{Diag}$ given in Eq.~(\ref{Gs}), and the relation between diagonal and angle dependent representations of the $Z_2\times Z_2$ symmetry in Eq.~(\ref{k4change}), allows for the explicit calculation of the Hermitian matrices $G_{i=1,2,3}$:
\begin{equation}
\begin{aligned}
G_1=&\left(
\begin{matrix}
 \left(G_1\right)_{11} & \left(G_1\right)_{12} & \left(G_1\right)_{13}  \\
 \left(G_1\right)_{12} ^* &\left(G_1\right)_{22} &\left(G_1\right)_{23}  \\
 \left(G_1\right)_{13}^*  &\left(G_1\right)_{23}^* & \left(G_1\right)_{33} \\
\end{matrix}
\right),\label{newGs}
G_2=\left(
\begin{matrix}
\left(G_2\right)_{11} & \left(G_2\right)_{12} &\left(G_2\right)_{13} \\
  \left(G_2\right)_{12}^* &\left(G_{2}\right)_{22} &  \left(G_{2}\right)_{23} \\
 \left(G_2\right)_{13}^*&  \left(G_{2}\right)_{23}^*& \left(G_{2}\right)_{33}
\end{matrix}
\right),\\~~~~~~&~~~
G_3=\left(
\begin{matrix}
 -c'_{13} & e^{-i \delta } s_{23} s'_{13} & - e^{-i \delta }c_{23} s'_{13} \\
 e^{i \delta } s_{23} s'_{13} & s_{23}^2 c'_{13}-c_{23}^2 & - c_{13}^2  s'_{23} \\
 -e^{i \delta } c_{23} s'_{13} & - c_{13}^2  s'_{23} & c_{23}^2 c'_{13}-s_{23}^2 \\
\end{matrix}
\right),
\end{aligned}
\end{equation}
where $s_{ij}=\sin(\theta_{ij})$, $c_{ij}=\cos(\theta_{ij})$, $s'_{ij}=\sin(2\theta_{ij})$, $c'_{ij}=\cos(2\theta_{ij})$, and
\begin{equation}\label{eq:Gelements}
\begin{aligned}
\left(G_1\right)_{11}&= c_{13}^2 c'_{12}-s_{13}^2,\\
%\left(G_1\right)_{12}&= 2 c_{12} c_{13} \left(-c_{23} s_{12}-e^{-i \delta } c_{12} s_{13} s_{23}\right),\\" Simplified" below
\left(G_1\right)_{12}&= -2 c_{12} c_{13} \left(c_{23} s_{12}+e^{-i \delta } c_{12} s_{13} s_{23}\right),\\
\left(G_1\right)_{13}&= 2 c_{12} c_{13} \left(e^{-i \delta } c_{12} c_{23} s_{13}-s_{12} s_{23}\right),\\
\left(G_1\right)_{22}&=-c_{23}^2 c'_{12}+s_{23}^2 \left(s_{13}^2 c'_{12}-c_{13}^2\right)+\cos (\delta ) s_{13} s'_{12} s'_{23}, \\
\left(G_1\right)_{23}&= c_{23} s_{23} c_{13}^2+s_{13} \left(i \sin (\delta )-\cos (\delta ) c'_{23}\right) s'_{12}+\frac{1}{4} c'_{12} \left(c'_{13}-3\right) s'_{23},\\
\left(G_1\right)_{33}&=\left(s_{13}^2 c'_{12}-c_{13}^2\right) c_{23}^2-s_{23}^2 c'_{12}-\cos (\delta ) s_{13} s'_{12} s'_{23}, \\
\left(G_2\right)_{11}&= -c'_{12} c_{13}^2-s_{13}^2,\\
\left(G_2\right)_{12}&= 2 c_{13} s_{12} \left(c_{12} c_{23}-e^{-i \delta } s_{12} s_{13} s_{23}\right), \\
\left(G_2\right)_{13}&= 2 c_{13} s_{12} \left(e^{-i \delta } c_{23} s_{12} s_{13}+c_{12} s_{23}\right),\\
\left(G_2\right)_{22}&= c'_{12} c_{23}^2-s_{23}^2 \left(c_{13}^2+s_{13}^2 c'_{12}\right)-\cos (\delta ) s_{13} s'_{12} s'_{23}, \\
%
%\left(G_2\right)_{23}&=e^{-i \delta } s_{13} s'_{12} c_{23}^2+\frac{1}{2} s_{23} \left(2 c_{13}^2-c'_{12} \left(c'_{13}-3\right)\right) c_{23}-2 e^{i \delta } c_{12} s_{12} s_{13} s_{23}^2,
%\\"Simplified" below
\left(G_2\right)_{23}&=e^{-i \delta } s_{13} s'_{12} c_{23}^2+\frac{1}{4} s'_{23} \left(2 c_{13}^2-c'_{12} \left(c'_{13}-3\right)\right) - e^{i \delta }  s'_{12} s_{13} s_{23}^2,
\\
\left(G_2\right)_{33}&=-c_{23}^2\left(c_{13}^2+s_{13}^2 c'_{12}\right) +s_{23}^2 c'_{12}+\cos (\delta ) s_{13} s'_{12} s'_{23} .
\end{aligned}
\end{equation}
Hence, for any choice of the mixing angles and Dirac phase, we can obtain a representation of the Klein group elements that are responsible for this prediction, cf.~Eq.~\eqref{newGs}.  Note that these results are independent of the Majorana phases, as can be seen from the fact that Eq.~(\ref{k4change}) is invariant under the transformation $U_\nu\rightarrow U_\nu P_{\rm Maj}$.  This feature is the reason why we choose to keep the neutrino mass eigenvalues complex and eliminate the Majorana phases in $U_\nu$.  We further comment that with the inclusion of the as yet unexplained matrix $P$ in Eq.~(\ref{eq:symunu}),  the eigenvector associated with the $+1$ eigenvalue of each of the above $G_i$ is exactly the $i^{th}$ column of $U_{\nu}$ in Eq.~(\ref{NewParam}), up to redefinition of Majorana phases.

To conclude this section, we will explicitly construct the form of the complex symmetric neutrino mass matrix that is invariant under the Klein group elements in Eq.~(\ref{newGs}) and diagonalized by Eq.~\eqref{eq:DiagMnu}, resulting in the complex neutrino mass eigenvalues $m_{1,2,3}$. This is of course easily done by rewriting Eq.~(\ref{eq:DiagMnu}) in the form
$M_\nu=U_\nu^* M_\nu^{\text{Diag}}U_\nu^\dagger$, and using the form of $U_{\nu}$ derived in Eq. \eqref{NewParam}.  The results are as follows:\footnote{Recall that the power of the residual Klein symmetry is that it fixes all of the mixing parameters except the Majorana phases. In contrast, if only a subgroup of this Klein symmetry is retained, there is an additional freedom to shift some of the mixing parameters. For example, if the subgroup retained is the $Z_2$ symmetry given by the identity and $G_2$, then the fact that $G_2^\text{Diag}$ is invariant under the transformation $G_2^\text{Diag} \rightarrow R_y^\dagger(\theta,\eta) G_2^\text{Diag} R_y (\theta,\eta)$ for any value of $\theta$, $\eta$ will mean that the corresponding mixings can shift nontrivially as functions of these parameters; see e.g.~Ref.~\cite{A4CP} for an explicit example of the type within $A_4$ models. Clearly, however, this freedom to shift the mixings is not permitted if the full Klein symmetry is intact.}
\begin{equation}\label{eq:Mnu}
\begin{aligned}
\left(M_{\nu}\right)_{11}&= c_{13}^2 m_2 s_{12}^2+c_{12}^2 c_{13}^2 m_1+e^{2 i \delta } m_3 s_{13}^2,\\
\left(M_{\nu}\right)_{12}&= c_{13} (c_{12} m_1 (-c_{23} s_{12}-c_{12} e^{-i \delta } s_{13} s_{23})+m_2 s_{12} (c_{12} c_{23}-e^{-i \delta } s_{12} s_{13} s_{23})+\\&+e^{i \delta } m_3 s_{13} s_{23}),\\
\left(M_{\nu}\right)_{13}&= c_{13} (-c_{23}  m_3 s_{13}e^{i \delta }+m_2 s_{12} (c_{12} s_{23}+c_{23} e^{-i \delta } s_{12} s_{13})+\\&+c_{12} m_1 (-s_{12} s_{23}+c_{12} c_{23} e^{-i \delta } s_{13})),\\
\left(M_{\nu}\right)_{22}&= m_1 (c_{23} s_{12}+c_{12} e^{-i \delta } s_{13} s_{23}){}^2+m_2 (c_{12} c_{23}-e^{-i \delta } s_{12} s_{13} s_{23}){}^2+c_{13}^2 m_3 s_{23}^2,\\
%
%\left(M_{\nu}\right)_{23}&= m_1 (-s_{12} s_{23}+c_{12} c_{23} e^{-i \delta } s_{13}) (-c_{23} s_{12}-c_{12} e^{-i \delta } s_{13} s_{23})+\\&  +m_2 (c_{12} s_{23}+c_{23} e^{-i \delta } s_{12} s_{13}) (c_{12} c_{23}-e^{-i \delta } s_{12} s_{13} s_{23})-c_{13}^2 c_{23} m_3 s_{23},\\"Simplified" below
\left(M_{\nu}\right)_{23}&= m_1 (s_{12} s_{23}-c_{12} c_{23} e^{-i \delta } s_{13}) (c_{23} s_{12}+c_{12} e^{-i \delta } s_{13} s_{23})+\\&  +m_2 (c_{12} s_{23}+c_{23} e^{-i \delta } s_{12} s_{13}) (c_{12} c_{23}-e^{-i \delta } s_{12} s_{13} s_{23})-c_{13}^2 c_{23} m_3 s_{23},\\
\left(M_{\nu}\right)_{33}&= m_2 (c_{12} s_{23}+c_{23} e^{-i \delta } s_{12} s_{13}){}^2+m_1 (-s_{12} s_{23}+c_{12} c_{23} e^{-i \delta } s_{13})^2+c_{13}^2 c_{23}^2 m_3.
\end{aligned}
\end{equation}
In the above, there is still the freedom to globally rephase masses and remove the phase from $m_1$ to render it real and positive. If this redefinition is performed, the number of phases is reduced to 3, {\it i.e.}, the 2 Majorana phases on $m_2$ and $m_3$ and the Dirac phase $\delta$.  Notice that the phase $\delta$ is identified easily as the Dirac CP-violating phase from our above parametrization, as can be checked from the calculation of the Jarlskog invariant \cite{Jarlskog} (see \cite{WBinv1,WBinv2,WBinv3,WBinv4,WBinv5,WBinv6,WBinv7,WBinv8,WBinv9,WBinv10,WBinv11,WBinv12,WBinv13} for other possible weak basis invariants).

%%%%%%%%%%%%%%%%%%%%%%%%%%%%%%%%%%

\section{General Residual CP Symmetries\label{sec:genCP}}

%%%%%%%%%%%%%%%%%%%%%%%%%%%%%%%%%%

Since we have constructed the form of the Klein symmetry for arbitrary mixing parameters and the most general neutrino mass matrix consistent with this symmetry, we now turn to the determination of the generalized CP transformations, $X$, that are consistent with this residual Klein symmetry.  This is done first by noting that the generalized CP symmetries, $X$,  must satisfy the low energy condition\cite{WBinv1}
\begin{align}
X^TM_\nu X&=M_\nu^*. \label{arbCP}
\end{align}
By rotating $M_\nu$ to the basis in which it is diagonal, it is straightforward to derive that 
\begin{equation}
\label{Xdiagrel}
X=U_{\nu}X^{\text{Diag}} U_{\nu}^{T},
\end{equation}
in which $X^{\text{Diag}}$ is given by
\begin{equation}\label{eq:diagX}
X^{\text{Diag}}=\left(
\begin{array}{ccc}
 \pm e^{i \alpha_ 1} & 0 & 0 \\
 0 & \pm e^{i \alpha_ 2} & 0 \\
 0 & 0 & \pm e^{i \alpha_ 3} \\
\end{array}
\right).
\end{equation}
where $\alpha_{1,2,3}$ are the Majorana phases as given in Eq.~(\ref{eq:DiagMnu}).  Equivalent expressions for $X$ have been given previously in the literature\cite{Feruglio:2012cw,Dingnew}, in which the Majorana phases that appear in Eq.~(\ref{eq:diagX}) have been absorbed into the definition of $U_\nu$. 

We pause here to comment on the role of overall phases of the $X$ transformation. If a global phase redefinition is performed on $M_{\nu}$, {\it i.e.}, $M_{\nu}\rightarrow e^{ i\theta}M_{\nu}$, then  $X\rightarrow e^{-i\theta} X$ to keep Eq.~(\ref{arbCP}) invariant.  This global rephasing of $M_\nu$ does not affect the mixing angles or the observable CP phases of the lepton sector, and hence it is unphysical.  However, the form of $X$ in Eq.~(\ref{eq:diagX}) is provided to keep the role of the individual Majorana phases in generalized CP transformations explicit, even though the freedom still exists to remove an overall phase. 

The Majorana assumption also introduces unphysical $\pm 1$  phases on each entry of the mass matrix, allowing  Eq.~\eqref{eq:diagX} to be rewritten in terms of the Klein symmetry elements as
\begin{equation}\label{eq:XiGi}
X_i^{\text{Diag}}=G_i^{\text{Diag}}\times\text{Diag}( e^{i \alpha_ 1} , e^{i \alpha_ 2} , e^{i \alpha_ 3} )~\text{for $i=0,1,2,3$,}
\end{equation}
in which the redundant negative determinant solutions have been discarded as previously was done when deriving the Klein symmetry elements.  Importantly, from the above equation it is possible to see that the $X_i^{\text{Diag}}$ represent a complexification of the $G_i^{\text{Diag}}$ of Eq.~(\ref{Gs}).  From Eq.~\eqref{klein} and Eq.~\eqref{eq:XiGi}, it is straightforward to deduce that
\begin{equation}\label{eq:XtoGdiag}
\begin{aligned}
&X_i^{\text{Diag}}(X_i^{\text{Diag}})^*=G_0=1~\text{for $i$ = 0, 1, 2, 3,}\\
&X_0^{\text{Diag}}(X_i^{\text{Diag}})^*=G_i^{\text{Diag}}~\text{for $i$ = 1, 2, 3,}\\%Case added for clarification.
&X_i^{\text{Diag}}(X_j^{\text{Diag}})^*=G_k^{\text{Diag}} ~\text{for $i\neq j\neq k\neq 0$.}
\end{aligned}
\end{equation}
Then, it becomes clear from Eq.~(\ref{eq:XtoGdiag}) that
\begin{equation}\label{eq:XtoG}
\begin{aligned}
&X_iX_i^*=G_0=1~\text{for $i$ = 0, 1, 2, 3,}\\
&X_0X_i^*=G_i~\text{for $i$ = 1, 2, 3,}\\%Case added for clarification.
&X_iX_j^*=G_k ~\text{for $i\neq j\neq k\neq 0$.}
\end{aligned}
\end{equation}
Hence,\footnote{Eq.~\eqref{eq:XtoG}, can also be obtained by demanding that two subsequent generalized CP transformations leave the mass matrix unchanged.  Namely, $X_j^{\dagger}X_i^TM_{\nu}X_iX_j^*=M_{\nu}$.}
\begin{equation}
\begin{aligned}
&X_iX_i^*=1~\text{for $i$ = 0, 1, 2, 3,}\\
&(X_0X_i^*)^2=1~\text{for $i=1,2,3,$}\\%Case added for clarification.
&(X_iX_j^*)^2=1~\text{for $i\neq j\neq 0$}
\end{aligned}
\end{equation}
 must always be conditions fulfilled by the generalized CP symmetries that are to be consistent with a residual Klein flavor symmetry group.

We comment here that if $X_iX_j^*=G'\neq G_k$ but instead some other symmetry transformation, then the residual symmetry group of the neutrino mass matrix will be larger than the original $Z_2\times Z_2$ Klein symmetry.  Recall that a Klein symmetry completely determines the lepton mixing matrix $U_{\rm MNSP}$ (up to charged lepton rephasing)  in the diagonal charged lepton basis, while leaving the complex neutrino masses as arbitrary parameters.   Therefore, a residual symmetry larger than a Klein symmetry introduces additional constraints on the neutrino mass matrix that lead to unphysical predictions for the masses and mixings, such as degeneracy of one more of the neutrino masses, cf. Eq.~(\ref{1b2}).  It is also worthwhile to remark here that if only part of the Klein symmetry is preserved, then mixing will not completely be determined and additional free parameters will enter the neutrino mixing matrix.  Either of these cases will not be considered in the approach outlined in this work because a phenomenologically viable framework in which the mixings are completely determined by the residual symmetry is the focus of this work (see Ref.~\cite{Dingnew} for an alternative treatment of a partially broken Klein symmetry consistent with a generalized CP symmetry).
 
In order to obtain more useful forms for the $X_i$, Eq.~(\ref{Xdiagrel}) should be expanded to reveal the most general CP symmetry consistent with an unbroken Klein symmetry.  Performing this expansion reveals the most general form of the symmetric $X_i=X_i^T$, as follows:
\begin{equation}\label{Xmatrix}
\begin{aligned}
X_{11}&=(-1)^a e^{i \alpha _1} c_{12}^2 c_{13}^2+ (-1)^b  e^{i \alpha _2} c_{13}^2 s_{12}^2+(-1)^c s_{13}^2 e^{i (\alpha _3-2 \delta) },\\
X_{12}&=(-1)^{a+1} e^{i \alpha _1} c_{12} c_{13} (c_{23} s_{12}+c_{12} s_{13} s_{23}e^{i \delta } )+ (-1)^b e^{i \alpha _2} c_{13} s_{12} (c_{12} c_{23}-\\&-s_{12} s_{13} s_{23}e^{i \delta } ) +(-1)^c c_{13} s_{13} s_{23} e^{i (\alpha _3-\delta) },\\
X_{13}&=(-1)^{a+1} e^{i \alpha _1} c_{12} c_{13} (s_{12} s_{23}-c_{12} c_{23} s_{13} e^{i \delta } )+(-1)^b e^{i \alpha _2}  c_{13} s_{12} (c_{12} s_{23}+\\&+c_{23} s_{12} s_{13} e^{i \delta } ) +(-1)^{c+1} c_{13} c_{23} s_{13} e^{i( \alpha _3-\delta) }, \\
X_{22}&=(-1)^{a} e^{i \alpha _1} (c_{23} s_{12}+c_{12} s_{13} s_{23}e^{i \delta } ){}^2+(-1)^b e^{i \alpha _2}  (c_{12} c_{23}-s_{12} s_{13} s_{23}e^{i \delta } ){}^2+\\&+(-1)^c e^{i \alpha _3} c_{13}^2 s_{23}^2,\\
X_{23}&=(-1)^{a} e^{i \alpha _1} (s_{12} s_{23}-c_{12} c_{23} s_{13}e^{i \delta } ) (c_{23} s_{12}+c_{12} s_{13} s_{23}e^{i \delta } )+\\
& +(-1)^b e^{i \alpha _2}  (c_{12} s_{23}+c_{23} s_{12} s_{13}e^{i \delta } ) (c_{12} c_{23}-s_{12} s_{13} s_{23}e^{i \delta } )+(-1)^{c+1} e^{i \alpha _3} c_{23} c_{13}^2 s_{23},\\
X_{33}&=(-1)^a e^{i \alpha _1} (s_{12} s_{23}-c_{12} c_{23} s_{13} e^{i \delta } ){}^2+(-1)^be^{i \alpha _2}  (c_{12} s_{23}+c_{23}s_{12} s_{13}e^{i \delta }  ){}^2+\\&+(-1)^{c} e^{i \alpha _3} c_{13}^2 c_{23}^2.
\end{aligned}
\end{equation}
%%%%%%%%%%%%%%%%%%%%%%%%%%%%%%%%%%%%%%%
in which we have introduced the parameters $a,b,c$ defined by the relations $(-1)^a=(G_i^{\text{Diag}})_{11}$, $(-1)^b=(G_i^{\text{Diag}})_{22}$, $(-1)^c=(G_i^{\text{Diag}})_{33}$, cf. Eq.\eqref{eq:XiGi}, to parametrize all four $X_i$ in one compact form that we have collectively denoted as $X$.

From the form of $X$ as given in Eq.~(\ref{Xmatrix}), it is straightforward to show that the $X_i$ and the $G_j$ always satisfy the known relation\cite{Feruglio:2012cw}:
\begin{equation}\label{eq:uncommute}
X_iG_j^*-G_jX_i=0 \text{~for $i,j=0,1,2,3$},
\end{equation}
for arbitrary values of CP phases.\footnote{Eq.~(\ref{eq:uncommute}) is directly attainable from simultaneously considering Eq.~(\ref{eq:groupinv}) and Eq.~(\ref{arbCP}).  It also serves as a check on the explicit form of $X$ in Eq.~(\ref{Xmatrix}), as well as the $G_i$ in Eqs.~(\ref{newGs})-(\ref{eq:Gelements}). }
However if $\delta=0$, $\pi$, the relationship between the residual flavor and CP symmetries in Eq.~(\ref{eq:uncommute}) reduces to
\begin{equation}
\label{eq:commute}
[X_i,G_j]_{\delta=0,\pi}=0 \text{~for $i,j=0,1,2,3$.}
\end{equation}
This can be easily understood by realizing that $G_i=G_i^*$ when $\delta=0,\pi,$ cf.~Eq.~(\ref{newGs}).
 Thus, even if the unbroken residual generalized CP symmetry $X_i$ has an order \textit{different} than $2$, it will always commute with the elements of the Klein group if the Dirac CP phase is trivial.  Therefore, in order to generate a nontrivial Dirac phase when the residual generalized CP and flavor symmetries commute in a specific model, the Dirac phase must have a separate origin, such as from charged lepton corrections as in Refs.~\cite{Ding:2014ssa,bimaxS4,tprime,Ding:2014ora}.

An additional bit of information concerning the relationship between the $X_i$ and $G_i$ results if  all of the Majorana phases are let to vanish.  If this occurs, then
\begin{equation}\label{eq:X-G}
(X_i-G_i)_{mn}\propto  (e^{2i\delta}-1)  \text{ for $i=0,1,2,3$.}
\end{equation}
Clearly, there is only equality between the $X_i$ and $G_i$ in Eq.~(\ref{eq:X-G}) if $\delta=0,\pi$ (without choosing special values for the angles, e.g.~a vanishing reactor mixing angle.)

Hence, by comparing Eqs.~(\ref{eq:uncommute})-(\ref{eq:X-G}), it is clear that if a nonvanishing Dirac phase in the neutrino sector is desired, then the residual flavor and CP symmetries must not commute and certainly not be equal.  However, if one wants such commutation and/or equality, then the nontrivial Dirac phase \textit{cannot} originate in the Klein symmetry itself but can be obtained from corrections to the charged lepton mixing, as previously mentioned (see text following Eq.~(\ref{eq:commute})).  Perhaps more importantly, the residual CP and flavor symmetries can commute with nontrivial Majorana phases, but they should not be equal unless vanishing Majorana (and Dirac) phases are desired. Therefore, it is clear that the key to understanding the Majorana phases in this framework is to understand the possible forms of the generalized CP symmetries, $X_i$.  

To this end, let us consider the situation in which $X_i$ is an element of a discrete symmetry group of finite order $m$.  In this case, there will exist an integer $n$ dividing $m$ such that
\begin{equation}\label{ordX}
X_i^n=1.
\end{equation}
Satisfying Eq.~(\ref{ordX}) will then impose nontrivial relations on the parameters of the theory.   For example, one feature that can result is that the Majorana phases are allowed to take only specific discrete values. As a trivial (but unrealistic) example of this phenomenon, we see that if the neutrino mass matrix was diagonal, the Majorana phases  $\alpha _{1}$, $\alpha _{2}$, and $\alpha _{3}$ must have values
\begin{equation}\label{discrete}
 \alpha _{1}=\frac{2\pi k_{1}}{n},
 \alpha _{2}=\frac{2\pi k_{2}}{n},
 \alpha _{3}=\frac{2\pi k_{3}}{n},
\end{equation}
where $k_1, k_{2},k_{3}=0,1\ldots,n-1$ to satisy Eq.~(\ref{ordX}). 
Another example  concerning the values of the Majorana phases can be deduced if the $X_i$ is assumed originate in $SU(3)$.  In this case, the determinant condition of $SU(3)$ implies that 
\begin{equation}\label{su3}
 \alpha_{1}+\alpha_{2}+\alpha_{3}=0~\text{mod}~2\pi.
\end{equation} 
Clearly, the above relation constrains the relative values of the Majorana phases, providing another prediction for the possible values of the Majorana phases.

Hence, making assumptions concerning the possible origins of the $X_i$ symmetries can lead to predictions concerning the values of the Majorana phases, as in~Eqs.~(\ref{discrete})-(\ref{su3}).  Such predictions can theoretically be probed at current and future neutrinoless double beta decay experiments and used to constrain the entries of $X_i^{\text{Diag}}$, assuming such a framework that is outlined in this work.  However, if the low energy mixing parameters are not taken as inputs for understanding the structure of the residual symmetry elements, it is possible to generate predictions for these parameters by constructing a model that spontaneously breaks a flavor group $G_f$ to $Z_2\times Z_2$ in the neutrino sector and to $Z_p$ ($p$ an integer) in the charged lepton sector,  while also breaking a consistently defined generalized CP symmety $H_{CP}$ to the $X_i$.  Of course, the scale of such breaking is presumed to be around the Grand Unified Theory (GUT) scale.  However, then the predictions for the mixing parameters can become subject to model-dependent corrections resulting from  charged lepton corrections, renormalization group evolution, and canonical normalization considerations.  One may expect such corrections to be subleading, as  renormalization group and canonical normalization effects are expected to be small in realistic models with hierarchical neutrino masses, and the charged lepton corrections in these models are typically at most Cabibbo-sized\cite{CLRGECN,genrge,CN}.\footnote{If $U_{\nu}$ is taken as a starting point for $U_{\text{MNSP}}$, relying only on $U_e$ for corrections to bring $U_{\nu}$ to the experimentally measured values, the resulting charged lepton corrections can be large\cite{CL}.}  Nonetheless, such corrections can in principle have nontrivial effects, particularly for the origin of Dirac-type CP violation in lepton mixing, since in many known examples the constraints due to the Klein symmetry as described above result in trivial values for $\delta_{\rm CP}$ at leading order.

As an example of the effects of such corrections, consider the $T'$ model of Ref.~\cite{tprime}.  In this model a $T'$ flavor symmetry is simultaneously implemented alongside its corresponding CP symmetry and broken to remnant symmetries that commute with each other at leading order.   Hence, this model's Dirac CP phase vanishes  at leading order, cf. Eq. \eqref{eq:commute}.  However by using model-dependent next-to-leading-order corrections to the charged lepton mass matrix, a non-diagonal form  for the charged lepton mixing is obtained.  This additional charged lepton rotation on $U_{\nu}$ provides the appropriate correction to render the CP phases as all nonzero.   In this example, it is the charged lepton corrections that  source a nontrivial $\delta_{\rm CP}$.  However, in principle it should be possible to construct scenarios in which leptonic Dirac CP violation results from other classes of corrections, such as from nontrivial Majorana phases at the high scale through renormalization group evolution, or through nontrivial Kahler metrics for the matter fields in supersymmetric theories.  A thorough classification of such corrections is beyond the scope of this current work.

%%%%%%%%%%%%%%%%%%%%%%%%%%%%%%%%%%%%%%

\section{Applications\label{sec:applications}}

%%%%%%%%%%%%%%%%%%%%%%%%%%%%%%%%%%%%%%%

We now turn to applying our method to specific examples, starting with the known examples in the literature of tribimaximal mixing \cite{TBmix} and bitrimaximal \cite{BTmixing} mixing, and then turning to a new analysis of golden ratio mixing \cite{GRmixing0,GRmixing1,A5GRmodels} models.   

Before discussing these specific examples, however, we note that a survey of the existing literature reveals that models with a preserved Klein symmetry based on an $A_4$ or $S_4$ flavor symmetry mostly predict $\theta_{13}=0$ and yield trivial predictions for the CP-violating phases when no corrections are considered\cite{A4CP,S4CP,Feruglio:2012cw,Feruglio:2013hia,Li:2013jya,bimaxS4}.  By considering corrections, e.g.~charged lepton corrections, a nontrivial value of the Dirac CP phase can be generated even though there is no contribution originating in the neutrino sector \cite{Ding:2014ssa,bimaxS4,tprime,Ding:2014ora}. Similarly, notice that if the Klein symmetry is broken or incomplete, then $U_\nu$ is not fully constrained, {\it i.e.},~it must contain at least one additional free parameter. This additional freedom can lead to nontrivial predictions \cite{Ding:2014hva,S4CP,tprime,Hagedorn:2014wha,Feruglio:2012cw,A4CP,Feruglio:2013hia,Ding:2014ssa,Li:2013jya,bimaxS4}.
However even though the Klein symmetry can be broken or incomplete, trivial CP phases can still result from the preservation of identical residual flavor and CP symmetry elements \cite{A4CP,Feruglio:2012cw,S4CP,Li:2013jya,bimaxS4,Ding:2014hva,Ding:2014ssa}. Thus, if nontrivial CP phases are desired, then the preserved flavor and CP symmetries must not be identical, e.g.~a trivial Dirac phase and a nontrivial Majorana phase prediction can result from a mismatch between the preserved flavor and generalized CP symmetries \cite{NandK,Hagedorn:2014wha,Ding:2014ora}.  Rather than providing a comprehensive review of all known examples, which is far beyond the scope of this work, our aim in this section is to explore a few of these simple scenarios in some detail, and to illustrate the utility of our approach.\\

\noindent $\bullet$ {\bf Tribimaximal mixing scenarios.}  
Perhaps the best example to begin the discussion of the applicability of the formalism presented here to the existing literature, is that of tribimaximal mixing\cite{TBmix}, for which 
\begin{equation}\label{eq:TBMang}
\theta_{12}^{\text{TBM}}=\tan^{-1}\left(\frac{1}{\sqrt{2}}\right),~~\theta_{23}^{\text{TBM}} = \frac \pi 4,~~ \theta_{13}^{\text{TBM}}=0,~~\delta^{\text{TBM}} =0.
\end{equation}
In this context, the MNSP matrix, cf. Eq.~\eqref{NewParam}, is clearly the tribimaximal (TBM) mixing matrix (up to charged lepton rephasing), which takes the following well-known form
\begin{align}
U^{\text{TBM}}=\left(\begin{matrix}\sqrt{\frac23}&\frac 1 {\sqrt{3}} & 0\\ - \frac 1 {\sqrt{6}}  & \frac 1 {\sqrt{3}}& \frac 1 {\sqrt{2}}\\ -\frac 1 {\sqrt{6}}& \frac 1 {\sqrt{3}}& -\frac 1 {\sqrt{2}}\end{matrix}\right).
\end{align}
Then by applying the values for the mixing angles given in Eq.~\eqref{eq:TBMang} to Eq.~\eqref{newGs}, the nontrivial Klein elements associated with tribimaximal mixing are
\begin{eqnarray}
\begin{aligned}
G_1^{\text{TBM}}=\frac 1 3\left(\begin{array}{ccc}
1&-2&-2\\ -2&-2&1 \\ -2 & 1 & -2\end{array}\right),
G_2^{\text{TBM}}=\frac 1 3&\left(\begin{array}{ccc}-1&2&2\\ 2&-1&2 \\ 2 & 2 & -1\end{array}\right),\\
 G_3^{\text{TBM}}=\left(\begin{array}{ccc}-1&0&0\\ 0&0&-1\\ 0&-1&0\end{array}\right),
\end{aligned}
\end{eqnarray}
which are the canonical $SU$, $S$, and $U$ elements of the Klein subgroup of $S_4$ associated with tribimaximal lepton mixing.   Furthermore, notice that the most general mass matrix associated with such a tribimaximal Klein symmetry can be found by using  the tribimaximal mixing angles as inputs for Eq. \eqref{eq:Mnu}:
\begin{align}\label{eq:TBMmass}
M_\nu^{\text{TBM}}&= \frac{1}{3}\left(
\begin{matrix}
 \left(2 m_1+m_2\right) &  \left(m_2-m_1\right) & \left(m_2-m_1\right) \\
  \left(m_2-m_1\right) & \frac{1}{2} \left(m_1+2 m_2+3 m_3\right) & \frac{1}{2} \left(m_1+2 m_2-3 m_3\right) \\
  \left(m_2-m_1\right) & \frac{1}{2} \left(m_1+2 m_2-3 m_3\right) & \frac{1}{2} \left(m_1+2 m_2+3 m_3\right) \\
\end{matrix}
\right).
\end{align}
Next, all possible generalized CP symmetries consistent with such a tribimaximal Klein symmetry can found by utilizing Eq. \eqref{Xmatrix}.  Doing this reveals the symmetry elements of $X$ to be 
\begin{equation}\label{eq:XTBM}
\begin{aligned}
X_{11}^{\text{TBM}}&=\frac{1}{3} \left(2 (-1)^a e^{i \alpha _1}+e^{i \alpha _2} (-1)^b\right),\\
X_{12}^{\text{TBM}}&=\frac{1}{3} \left((-1)^{a+1} e^{i \alpha _1}+e^{i \alpha _2} (-1)^b\right),\\
X_{13}^{\text{TBM}}&=\frac{1}{3} \left((-1)^{a+1} e^{i \alpha _1}+e^{i \alpha _2} (-1)^b\right),\\
X_{22}^{\text{TBM}}&= \frac{1}{6} \left((-1)^a e^{i \alpha _1}+2 e^{i \alpha _2} (-1)^b+3 e^{i \alpha _3} (-1)^c\right),\\
X_{23}^{\text{TBM}}&= \frac{1}{6} \left((-1)^a e^{i \alpha _1}+2 e^{i \alpha _2} (-1)^b-3 e^{i \alpha _3} (-1)^c\right),\\
X_{33}^{\text{TBM}}&=\frac{1}{6} \left((-1)^a e^{i \alpha _1}+2 e^{i \alpha _2} (-1)^b+3 e^{i \alpha _3} (-1)^c\right),
\end{aligned}
\end{equation}
where the parameters $a$, $b$, and $c$ have been defined below Eq. \eqref{Xmatrix}.  It is trivial to show that the above solutions for $X^{\text{TBM}}$ and the mass matrix of Eq.\eqref{eq:TBMmass} satisfy the low energy condition for generalized CP symmetries, {\it i.e.},~Eq. \eqref{arbCP}.  Additionally, notice that the generalized CP symmetries given in Eq. \eqref{eq:XTBM} are functions of the three Majorana phases.  When working in a top-down approach these parameters will be given by elements of the automorphism group of the flavor symmetry group\cite{lindner}; however, in the bottom-up approach taken in this work, these Majorana phases become inputs for the $X_i$.  Therefore, let us here \textit{assume} that the Majorana phases have been measured and are consistent with $0$ or $\pi$.  Applying these trivial values to Eq. \eqref{eq:XTBM}, reveals that the generalized CP symmetries $X_i^{\text{TBM}}$ to be identical to the Klein elements $1$, $S$, $U$, and $SU$, members of the automorphism group of $S_4$ \cite{S4CP,Feruglio:2012cw,Feruglio:2013hia}.  On the other hand, it is more likely that the measured Majorana phases will  not be trivial.  Even if this is the case, $X^{\text{TBM}}$ will still give the allowed CP symmetries consistent with tribimaximal mixing.  However, $X^{\text{TBM}}$ will no longer be an element of the automorphism group of $S_4$, but \textit{perhaps} of the automorphism group of some other flavor symmetry group that can also predict tribimaximal neutrino mixing (see e.g.~Refs.~\cite{scans0,scans1} for examples of scans over possible flavor symmetry groups that contain  viable flavor subgroups).\\  

\noindent $\bullet$ {\bf Bitrimaximal mixing scenarios.}  
Let us now consider an example from the literature that predicts a nonzero reactor mixing angle and a non-maximal atmospheric mixing angle at leading order, e.g.~bitrimaximal mixing\cite{BTmixing}.  The bitrimaximal mixing pattern is 
\begin{equation}\label{eq:BTMang}
\theta_{12}^{\text{BTM}}=\theta_{23}^{\text{BTM}}=\tan^{-1}(\sqrt{3}-1),\theta_{13}^{\text{BTM}}=\sin^{-1}(\frac{1}{6}(3-\sqrt{3})), \delta^{\text{BTM}}=0.
\end{equation}
Using these values as inputs, the bitrimaximal MNSP matrix can be shown to have the form
\begin{align}
U^{\text{BTM}}&=\left(
\begin{matrix}
 \frac{1}{6} \left(3+\sqrt{3}\right) & \frac{1}{\sqrt{3}} & \frac{1}{6} \left(3-\sqrt{3}\right) \\
 -\frac{1}{\sqrt{3}} & \frac{1}{\sqrt{3}} & \frac{1}{\sqrt{3}} \\
 \frac{1}{6} \left(-3+\sqrt{3}\right) & \frac{1}{\sqrt{3}} & \frac{1}{6} \left(-3-\sqrt{3}\right) \\
\end{matrix}
\right).
\end{align}
Applying the values for the mixing angles given in Eq.~\eqref{eq:BTMang} to Eq.~\eqref{newGs}, reveals the nontrivial Klein elements associated with bitrimaximal mixing to be
\begin{equation}
\begin{aligned}
G_1^{\text{BTM}}&=\left(
\begin{matrix}
 \frac{1}{\sqrt{3}}-\frac{1}{3} & -\frac{1}{3}-\frac{1}{\sqrt{3}} & -\frac{1}{3} \\
 -\frac{1}{3}-\frac{1}{\sqrt{3}} & -\frac{1}{3} & \frac{1}{\sqrt{3}}-\frac{1}{3} \\
 -\frac{1}{3} & \frac{1}{\sqrt{3}}-\frac{1}{3} & -\frac{1}{3}-\frac{1}{\sqrt{3}} \\
\end{matrix}
\right),  G_2^{\text{BTM}} = \frac{1}{3}\left(
\begin{matrix}
 -1 & 2 & 2\\
2& -1& 2 \\
 2 &2 & -1 \\
\end{matrix}
\right),\\&~~~~~ G_3^{\text{BTM}}= \left(
\begin{matrix}
 -\frac{1}{3}-\frac{1}{\sqrt{3}} & \frac{1}{\sqrt{3}}-\frac{1}{3} & -\frac{1}{3} \\
 \frac{1}{\sqrt{3}}-\frac{1}{3} & -\frac{1}{3} & -\frac{1}{3}-\frac{1}{\sqrt{3}} \\
 -\frac{1}{3} & -\frac{1}{3}-\frac{1}{\sqrt{3}} & \frac{1}{\sqrt{3}}-\frac{1}{3} \\
\end{matrix}
\right).
\end{aligned}
\end{equation}
The above bitrimaximal Klein elements match those  found in the literature\cite{BTmixing}.
The most general symmetric mass matrix invariant under bitrimaximal symmetry is also easily found by using Eq. \eqref{eq:Mnu} with the angles given in Eq. \eqref{eq:BTMang} as input.  This matrix is
\begin{equation}
\begin{aligned}
(M_\nu^{\text{BTM}})_{11} &= \frac{1}{6} ((2+\sqrt{3}) m_1+2 m_2-(-2+\sqrt{3}) m_3),\\
(M_\nu^{\text{BTM}})_{12} &=  \frac{1}{6} (-(1+\sqrt{3}) m_1+2 m_2+(-1+\sqrt{3}) m_3),\\
(M_\nu^{\text{BTM}})_{13} &= \frac{1}{6} (-m_1+2 m_2-m_3),\\
(M_\nu^{\text{BTM}})_{22} & = \frac{1}{3} (m_1+m_2+m_3),\\
(M_\nu^{\text{BTM}})_{23} &=  \frac{1}{6} ((-1+\sqrt{3}) m_1+2 m_2-(1+\sqrt{3}) m_3),\\
(M_\nu^{BTM})_{33} &=  \frac{1}{6} (-(-2+\sqrt{3}) m_1+2 m_2+(2+\sqrt{3}) m_3).
\end{aligned}
\end{equation}
Finally, it is straightforward to calculate the bitrimaximal generalized CP symmetries using Eq. \eqref{Xmatrix}  and the bitrimaximal angle values of Eq. \eqref{eq:BTMang}.  For this example, $X$ is given by
\begin{equation}
\begin{aligned}
X^{\text{BTM}}_{11}&=  \frac{1}{6} \left((-1)^{c+1} e^{i \alpha _3} \left(-2+\sqrt{3}\right)+(-1)^a \left(2+\sqrt{3}\right) e^{i \alpha _1}+2 (-1)^b e^{i \alpha _2}\right), \\ 
X^{\text{BTM}}_{12} & = \frac{1}{6} \left((-1)^c e^{i \alpha _3} \left(-1+\sqrt{3}\right)+(-1)^{a+1} \left(1+\sqrt{3}\right) e^{i \alpha _1}+2 (-1)^b e^{i \alpha _2}\right),\\
X^{\text{BTM}}_{13} & =\frac{1}{6} \left((-1)^{a+1} e^{i \alpha _1}+2 (-1)^b e^{i \alpha _2}+(-1)^{c+1} e^{i \alpha _3}\right), \\
X^{\text{BTM}}_{22} & =\frac{1}{3} \left((-1)^a e^{i \alpha _1}+(-1)^b e^{i \alpha _2}+(-1)^c e^{i \alpha _3}\right),\\
X^{\text{BTM}}_{23} & = \frac{1}{6} \left((-1)^a e^{i \alpha _1} \left(-1+\sqrt{3}\right)+2 (-1)^b e^{i \alpha _2}+(-1)^{c+1} \left(1+\sqrt{3}\right) e^{i \alpha _3}\right), \\
X^{\text{BTM}}_{33} & =\frac{1}{6} \left((-1)^{a+1} e^{i \alpha _1} \left(-2+\sqrt{3}\right)+2 (-1)^b e^{i \alpha _2}+(-1)^c \left(2+\sqrt{3}\right) e^{i \alpha _3}\right). 
\end{aligned}
\end{equation}

As in the previous tribimaximal mixing case, if the Majorana phases are taken to be trivial, then these generalized CP transformations are identical to the Klein symmetry elements associated with bitrimaximal mixing and can be consistent with a $\Delta(96)$ flavor symmetry, and its  corresponding automorphism group\cite{NandK,Ding:2014ssa} that makes up the elements of the bitrimaximal generalized CP symmetry.  However, if the Majorana phases are not taken to be trivial but instead (for example) $\alpha_1=\alpha_3=\frac \pi 6$,  $\alpha_2=-\frac \pi 3$ with $a=1,b=0,c=1$, then $X^{\text{BTM}}\rightarrow X^{\text{BTM}}_2$ becomes an order 4 element of the automorphism group of $\Delta(96)$ \cite{NandK,Ding:2014ssa}, after global phase redefinition.  Of course, all of the other bitrimaximal generalized CP solutions consistent with the automorphism group of $\Delta(96)$ can be found in this manner as well as additional possibilities for $X^{\text{BTM}}$ that could exist if the bitrimaximal Klein symmetry was taken as the subgroup of a larger flavor symmetry group than $\Delta(96)$.  However, we note that even though this second case predicts nonzero Majorana phases, it still requires a vanishing Dirac CP phase to be consistent with the  (real) bitrimaximal Klein symmetry.  \\

 \noindent $\bullet$ {\bf Golden ratio mixing.}  Until this point, we have focused in this section on demonstrating that the formalism here is consistent with existing works in the literature.  However, the remainder of this section is devoted to making predictions on the possible generalized CP symmetries using another well-known mixing pattern, {\it i.e.}, the specific golden ratio mixing pattern discussed in Refs.~\cite{GRmixing0,GRmixing1,A5GRmodels}, which is often called the ``GR1" pattern in the literature.  

The GR1 mixing pattern has a vanishing reactor angle, a maximal atmospheric angle, and a solar mixing angle related to the golden ratio, $\phi$, of Grecian lore, as follows:
\begin{equation}\label{eq:GRang}
\theta_{12}^{\text{GR1}}=\tan^{-1}\left(\frac{1}{\phi}\right),~~\theta_{23}^{\text{GR1}} = \frac \pi 4,~~ \theta_{13}^{\text{GR1}}=0,~~\delta^{\text{GR1}} =0, 
\end{equation}
in which the golden ratio $ \phi=(1+\sqrt{5})/2$. Hence the lepton mixing matrix associated with this mixing pattern is given in the diagonal charged lepton basis by
\begin{eqnarray}
U^{\text{GR1}}=\left ( \begin{array}{ccc} 
\sqrt{\frac{\phi}{\sqrt{5}}} & \sqrt{\frac{1}{\sqrt{5}\phi}} & 0\\ -\frac{1}{\sqrt{2}}\sqrt{\frac{1}{\sqrt{5}\phi}} & \frac{1}{\sqrt{2}}\sqrt{\frac{\phi}{\sqrt{5}} }& \frac{1}{\sqrt{2}}\\
-\frac{1}{\sqrt{2}}\sqrt{\frac{1}{\sqrt{5}\phi}} & \frac{1}{\sqrt{2}}\sqrt{\frac{\phi}{\sqrt{5}} }& -\frac{1}{\sqrt{2}}\\
\end{array} \right ).
\end{eqnarray}
As before the  Klein elements responsible for golden ratio lepton mixing can be calculated, cf. Eq. \eqref{newGs}, to be
\begin{equation}\label{eq:GRKlein}
\begin{aligned}
G_1^{\text{GR1}}=\frac{1}{\sqrt{5}}\left(
\begin{array}{ccc}
 1 & -\sqrt{2} & -\sqrt{2} \\
 -\sqrt{2} & -\phi  & \phi -1 \\
 -\sqrt{2} & \phi -1 & -\phi  \\
\end{array}
\right),G_2^{\text{GR1}}=\frac{1}{\sqrt{5}}&\left(
\begin{array}{ccc}
 -1 & \sqrt{2} & \sqrt{2} \\
 \sqrt{2} & 1-\phi  & \phi  \\
 \sqrt{2} & \phi  & 1-\phi  
\end{array}\right),\\
G_3^{\text{GR1}}=\left(
\begin{array}{ccc}
 -1 & 0 & 0 \\
 0 & 0 & -1 \\
 0 & -1 & 0 \\
\end{array}
\right),&~~~~~
\end{aligned}
\end{equation}
matching results in Ref.~\cite{A5GRmodels}.
As before in the previous examples, it is trivial to find the most general symmetric mass matrix invariant under this golden ratio Klein symmetry by using Eq. \eqref{eq:Mnu} with the angles given in Eq. \eqref{eq:GRang} as input.  This ``golden'' mass matrix is
\begin{eqnarray}
M_{\nu}^{\text{GR1}}=\frac{1}{\sqrt{5}}\left(
\begin{array}{ccc}
 \frac{m_1 \phi ^2+m_2}{\phi } & \frac{m_2-m_1}{\sqrt{2}} & \frac{m_2-m_1}{\sqrt{2}} \\
 \frac{m_2-m_1}{\sqrt{2}} & \frac{\left(m_2+m_3\right) \phi ^2+m_1+m_3}{2 \phi } & \frac{m_2 \phi ^2-\sqrt{5} m_3 \phi +m_1}{2 \phi } \\
 \frac{m_2-m_1}{\sqrt{2}} & \frac{m_2 \phi ^2-\sqrt{5} m_3 \phi +m_1}{2 \phi } & \frac{\left(m_2+m_3\right) \phi ^2+m_1+m_3}{2 \phi } \\
\end{array}
\right).
\end{eqnarray}
Finally, it is straightforward to calculate the symmetric golden ratio generalized CP symmetries using Eq. \eqref{Xmatrix}  and the golden ratio mixing values of Eq. \eqref{eq:GRang}.   They are
\begin{equation}\label{eq:XGR}
\begin{aligned}
X^{\text{GR1}}_{11}&=\frac{(-1)^a e^{i \text{$\alpha_1$}} \phi ^2+e^{i \text{$\alpha_2$}} (-1)^b}{\sqrt{5} \phi },\\ 
%X^{\text{GR1}}_{12}&=\frac{(-1)^{a+1} e^{i \text{$\alpha_1$}} \phi +e^{i \text{$\alpha_2$}} (-1)^b \phi }{\sqrt{10} \phi },\\Cancelled extra phi
X^{\text{GR1}}_{12}&=\frac{(-1)^{a+1} e^{i \text{$\alpha_1$}} +e^{i \text{$\alpha_2$}} (-1)^b  }{\sqrt{10} },\\
%X^{\text{GR1}}_{13}&=\frac{(-1)^{a+1} e^{i \text{$\alpha_1$}} \phi +e^{i \text{$\alpha_2$}} (-1)^b \phi }{\sqrt{10} \phi },\\Cancelled extra phi
X^{\text{GR1}}_{13}&=\frac{(-1)^{a+1} e^{i \text{$\alpha_1$}} +e^{i \text{$\alpha_2$}} (-1)^b  }{\sqrt{10}  },\\
X^{\text{GR1}}_{22}&=\frac{(-1)^a e^{i \text{$\alpha_1$}}+e^{i \text{$\alpha_2$}} (-1)^b \phi ^2+\sqrt{5} e^{i \text{$\alpha_3$}} (-1)^c \phi }{2 \sqrt{5}
   \phi },\\
X^{\text{GR1}}_{23}&=\frac{(-1)^a e^{i \text{$\alpha_1$}}+e^{i \text{$\alpha_2$}} (-1)^b \phi ^2+\sqrt{5} e^{i \text{$\alpha_3$}} (-1)^{c+1} \phi }{2
   \sqrt{5} \phi },\\
X^{\text{GR1}}_{33}&=\frac{(-1)^a e^{i \text{$\alpha_1$}}+e^{i \text{$\alpha_2$}} (-1)^b \phi ^2+\sqrt{5} e^{i \text{$\alpha_3$}} (-1)^c \phi }{2 \sqrt{5}
   \phi }.
\end{aligned}
\end{equation}
Clearly from the above equation, when all of the Majorana phases vanish, then the nontrivial $X_{i=1,2,3}^{\text{GR1}}$ are equivalent to the Klein symmetry elements in Eq.~\eqref{eq:GRKlein}.  Said again explicitly, when $\alpha_{1,2,3}=0$ in Eq. \eqref{eq:XGR}, then the golden ratio Klein elements can be obtained by applying the relevant values for $a,b,c$ to Eq.~\eqref{eq:XGR}.  For example (in the limit that all Majorana phases vanish), using $a=0,b=1,c=1$ yields exactly $G_1^{\text{GR1}}$; using $a=1,b=0,c=1$ yields exactly $G_2^{\text{GR1}}$; and  using $a=1,b=1,c=0$ yields exactly $G_3^{\text{GR1}}$. 

 Taking this logic one step further, it is possible to obtain the golden ratio Klein subgroup from an $A_5$ flavor symmetry group\cite{A5GRmodels}.  However, it need not be $A_5$.  In fact, the golden ratio prediction could come from a larger, different group all together.  In either case, any phenomenologically viable model predicting golden ratio mixing will have a set of ``golden" generalized CP symmetries  obtainable from Eq. \eqref{eq:XGR}, demonstrating the power of the bottom-up approach contained in this work. 

We close this section by noting that each of these examples predict trivial Dirac CP violation (here $\delta =0$).  Certainly one possibility is that leading order mixing patterns of this type, once corrected appropriately so as to be phenomenologically viable, are an appropriate starting point for flavor model-building, which in turn has been quite fruitful in terms of considering possible discrete non-Abelian family symmetry groups that can lead to these predictions.  However, one utility of the bottom-up approach advocated here is that by considering the mixing parameters as inputs and constructing the Klein elements and the generalized CP transformations, we can identify specific representations that may help in elucidating more general possibilities for the underlying family symmetry group.  In this way, this approach can provide a helpful guideline for future flavor model-building.

%%%%%%%%%%%%%%%%%%%%%%%%%%%%%%%%%%%%%%

\section{Conclusions\label{sec:conclusion}}

%%%%%%%%%%%%%%%%%%%%%%%%%%%%%%%%%%%%%%

If neutrinos are Majorana particles, the possibility exists that there is a high scale flavor symmetry that is spontaneously broken to a residual Klein symmetry at low energies.  If such a Klein symmetry is preserved, then it completely determines the mixing angles of the neutrino sector and it also produces specific relations between the entries of the neutrino mass matrix, $M_{\nu}$, but it is unable to provide predictions for the Majorana phases of the neutrinos.  In order to produce such predictions, a popular method  is to implement a generalized CP symmetry consistently alongside of the flavor symmetry and spontaneously break both symmetries within specific, top-down scenarios.
Within this top-down approach the exact roles that the generalized CP symmetry and flavor symmetry play in predicting the lepton mixing parameters are not easy to clarify, and can appear quite model-dependent.

In this work, we have constructed a bottom-up approach that clarifies the roles of the flavor and generalized CP symmetries in lepton mixing, by expressing the residual, unbroken Klein and generalized CP symmetries in terms of the lepton mixing parameters.   By doing this, it becomes clearly seen that  a nonzero prediction for ``Dirac''-type CP violation in the neutrino sector must originate in the Klein symmetry unless model-dependent corrections are utilized.  Perhaps more importantly is that, by keeping the neutrino masses as complex, we see that the generalized CP symmetries are the harbingers for Majorana phase predictions.  We have shown that this formalism  is able to reproduce results in the literature based on tribimaximal and bitrimaximal neutrino mixing and have demonstrated its power by predicting the generalized CP symmetries consistent with a certain type of golden ratio mixing (the GR1 pattern).  This method can serve as guidance for future model-building by identifying the appropriate symmetries and breaking patterns which need to occur to generate desired predictions for the lepton mixing angles and CP phases.

\section*{Acknowledgements}
L.E., T.G., and A.S.~would like to acknowledge useful discussions with S. Petcov, I. Girardi, and A.~Titov. L.E.~and T.G.~are supported by the U.~S.~Department of Energy under the contract DE-FG-02-95ER40896.  L.E.~acknowledges the support and hospitality of the Enrico Fermi Institute at the University of Chicago.  
A.S.~acknowledges support from the research grant  2012CPPYP7 ({\it  Theoretical Astroparticle Physics}) under the program  PRIN 2012 funded by the Italian Ministry of Education, University and Research (MIUR), as well as partial support from the European Union FP7 ITN INVISIBLES (Marie Curie Actions, PITN-GA-2011-289442-INVISIBLES).

%\newpage


\begin{thebibliography}{99}
\bibitem{dayabay}
%\bibitem{t13}
%\cite{An:2012eh}
%\bibitem{An:2012eh} 
  F.~P.~An {\it et al.}  [DAYA-BAY Collaboration],
  %``Observation of electron-antineutrino disappearance at Daya Bay,''
  Phys.\ Rev.\ Lett.\  {\bf 108}, 171803 (2012)
  [arXiv:1203.1669 [hep-ex]].
  %%CITATION = ARXIV:1203.1669;%%

\bibitem{reno}
%\cite{Ahn:2012nd}
%\bibitem{Ahn:2012nd} 
  J.~K.~Ahn {\it et al.}  [RENO Collaboration],
  %``Observation of Reactor Electron Antineutrino Disappearance in the RENO Experiment,''
  Phys.\ Rev.\ Lett.\  {\bf 108}, 191802 (2012)
  [arXiv:1204.0626 [hep-ex]].
  %%CITATION = ARXIV:1204.0626;%%

\bibitem{doublechooz}
%\cite{Abe:2014bwa}
%\bibitem{Abe:2014bwa} 
  Y.~Abe {\it et al.}  [Double Chooz Collaboration],
  %``Improved measurements of the neutrino mixing angle $\theta_{13}$ with the Double Chooz detector,''
  JHEP {\bf 1410}, 86 (2014)
  [arXiv:1406.7763 [hep-ex]].
  %%CITATION = ARXIV:1406.7763;%%

\bibitem{pdg}
K.A. Olive et al. (Particle Data Group), Chin. Phys. C, 38, 090001 (2014). 


\bibitem{global}
%\cite{Capozzi:2013csa}
%\bibitem{Capozzi:2013csa} 
  F.~Capozzi, G.~L.~Fogli, E.~Lisi, A.~Marrone, D.~Montanino and A.~Palazzo,
  %``Status of three-neutrino oscillation parameters, circa 2013,''
  Phys.\ Rev.\ D {\bf 89}, 093018 (2014)
  [arXiv:1312.2878 [hep-ph]];
  %%CITATION = ARXIV:1312.2878;%%
%\cite{Gonzalez-Garcia:2014bfa}
%\bibitem{Gonzalez-Garcia:2014bfa} 
  M.~C.~Gonzalez-Garcia, M.~Maltoni and T.~Schwetz,
  %``Updated fit to three neutrino mixing: status of leptonic CP violation,''
  arXiv:1409.5439 [hep-ph];
  %%CITATION = ARXIV:1409.5439;%%
%\cite{Forero:2014bxa}
%\bibitem{Forero:2014bxa} 
  D.~V.~Forero, M.~Tortola and J.~W.~F.~Valle,
  %``Neutrino oscillations refitted,''
  arXiv:1405.7540 [hep-ph].
  %%CITATION = ARXIV:1405.7540;%%



\bibitem{lindner}
%\cite{Holthausen:2012dk}
%\bibitem{Holthausen:2012dk} 
  M.~Holthausen, M.~Lindner and M.~A.~Schmidt,
  %``CP and Discrete Flavour Symmetries,''
  JHEP {\bf 1304}, 122 (2013)
  [arXiv:1211.6953 [hep-ph]].
  %%CITATION = ARXIV:1211.6953;%%

%%%%%%%%%%%%%%%%%%%%%%BEGIN GENCP MODEL REFS
%%%%A4
\bibitem{A4CP}
%\cite{Ding:2013bpa}
%\bibitem{Ding:2013bpa} 
  G.~J.~Ding, S.~F.~King and A.~J.~Stuart,
  %``Generalised CP and $A_4$ Family Symmetry,''
  JHEP {\bf 1312}, 006 (2013)
  [arXiv:1307.4212 [hep-ph]].
  %%CITATION = ARXIV:1307.4212;%%

%%%Delta(3n^2)

%\cite{Ding:2014hva}
\bibitem{Ding:2014hva} 
  G.~J.~Ding and Y.~L.~Zhou,
  %``Lepton mixing parameters from $\Delta(48)$ family symmetry and generalised CP,''
  JHEP {\bf 1406}, 023 (2014)
  [arXiv:1404.0592 [hep-ph]].
  %%CITATION = ARXIV:1404.0592;%%

%\cite{Hagedorn:2014wha}
\bibitem{Hagedorn:2014wha} 
  C.~Hagedorn, A.~Meroni and E.~Molinaro,
  %``Lepton Mixing from $\Delta$ (3 $n^2$) and $\Delta$ (6 $n^2$) and CP,''
  arXiv:1408.7118 [hep-ph].
  %%CITATION = ARXIV:1408.7118;%%

%%%%%%%%%%%%%S4


%\cite{Feruglio:2012cw}S4
\bibitem{Feruglio:2012cw} 
  F.~Feruglio, C.~Hagedorn and R.~Ziegler,
  %``Lepton Mixing Parameters from Discrete and CP Symmetries,''
  JHEP {\bf 1307}, 027 (2013)
  [arXiv:1211.5560 [hep-ph]].
  %%CITATION = ARXIV:1211.5560;%%

\bibitem{S4CP}
%\cite{Ding:2013hpa}
%\bibitem{Ding:2013hpa} 
  G.~J.~Ding, S.~F.~King, C.~Luhn and A.~J.~Stuart,
  %``Spontaneous CP violation from vacuum alignment in $S_4$ models of leptons,''
  JHEP {\bf 1305}, 084 (2013)
  [arXiv:1303.6180 [hep-ph]].
  %%CITATION = ARXIV:1303.6180;%%

%\cite{Feruglio:2013hia}
\bibitem{Feruglio:2013hia} 
  F.~Feruglio, C.~Hagedorn and R.~Ziegler,
  %``A realistic pattern of lepton mixing and masses from $S_4$ and CP,''
  Eur.\ Phys.\ J.\ C {\bf 74}, 2753 (2014)
  [arXiv:1303.7178 [hep-ph]].
  %%CITATION = ARXIV:1303.7178;

%\cite{Li:2013jya}
\bibitem{Li:2013jya} 
  C.~C.~Li and G.~J.~Ding,
  %``Generalised CP and trimaximal $TM_1$ lepton mixing in $S_4$ family symmetry,''
  Nucl.\ Phys.\ B {\bf 881}, 206 (2014)
  [arXiv:1312.4401 [hep-ph]].
  %%CITATION = ARXIV:1312.4401;%%

\bibitem{bimaxS4}
%\cite{Li:2014eia}
%\bibitem{Li:2014eia} 
  C.~C.~Li and G.~J.~Ding,
  %``Deviation from Bimaximal Mixing and Leptonic CP Phases in $S_4$ Family Symmetry and Generalized CP,''
  arXiv:1408.0785 [hep-ph].
  %%CITATION = ARXIV:1408.0785;%%


%%%%%%%%%Delta(6n^2)
\bibitem{NandK}
%\cite{King:2014rwa}
%\bibitem{King:2014rwa} 
  S.~F.~King and T.~Neder,
  %``Lepton mixing predictions including Majorana phases from Δ(6n$^2$) flavour symmetry and generalised CP,''
  Phys.\ Lett.\ B {\bf 736}, 308 (2014)
  [arXiv:1403.1758 [hep-ph]].
  %%CITATION = ARXIV:1403.1758;%%

%\cite{Ding:2014ssa}
\bibitem{Ding:2014ssa} 
  G.~J.~Ding and S.~F.~King,
  %``Generalised CP and $\Delta (96)$ Family Symmetry,''
  Phys.\ Rev.\ D {\bf 89}, 093020 (2014)
  [arXiv:1403.5846 [hep-ph]].
  %%CITATION = ARXIV:1403.5846;%%

%\cite{Ding:2014ora}
\bibitem{Ding:2014ora} 
  G.~J.~Ding, S.~F.~King and T.~Neder,
  %``Generalised CP and $\Delta (6n^2)$ Family Symmetry in Semi-Direct Models of Leptons,''
  arXiv:1409.8005 [hep-ph].

%%%%%%%%Sigma(nphi)

\bibitem{sigmagroups}
%\cite{Hagedorn:2013nra}
%\bibitem{Hagedorn:2013nra} 
  C.~Hagedorn, A.~Meroni and L.~Vitale,
  %``Mixing patterns from the groups $\Sigma(n\phi)$,''
  J.\ Phys.\ A {\bf 47}, 055201 (2014)
  [arXiv:1307.5308 [hep-ph]].
  %%CITATION = ARXIV:1307.5308;%

%%%%%%%%%%%%%%%%%%%T'
\bibitem{tprime}
%\cite{Girardi:2013sza}
%\bibitem{Girardi:2013sza} 
  I.~Girardi, A.~Meroni, S.~T.~Petcov and M.~Spinrath,
  %``Generalised geometrical CP violation in a T' lepton flavour model,''
  JHEP {\bf 1402}, 050 (2014)
  [arXiv:1312.1966 [hep-ph]].
  %%CITATION = ARXIV:1312.1966;%%

%%%%%%%%%%%%%%%%%%%%%%%%%%


%%%%%%%%%%%%%%%%%%%%CP and Z2 Refs

\bibitem{Z2CP1}
%\cite{Ge:2011ih}
%\bibitem{Ge:2011ih} 
  S.~F.~Ge, D.~A.~Dicus and W.~W.~Repko,
  %``Z_2 Symmetry Prediction for the Leptonic Dirac CP Phase,''
  Phys.\ Lett.\ B {\bf 702}, 220 (2011)
  [arXiv:1104.0602 [hep-ph]].
  %%CITATION = ARXIV:1104.0602;%%
%\cite{Ge:2011qn}

\bibitem{Z2CP2}
%\bibitem{Ge:2011qn} 
  S.~F.~Ge, D.~A.~Dicus and W.~W.~Repko,
  %``Residual Symmetries for Neutrino Mixing with a Large $\theta_{13}$ and Nearly Maximal $\delta_D$,''
  Phys.\ Rev.\ Lett.\  {\bf 108}, 041801 (2012)
  [arXiv:1108.0964 [hep-ph]].
  %%CITATION = ARXIV:1108.0964;%%

\bibitem{Z2CP3}
%\cite{Hernandez:2012ra}
%\bibitem{Hernandez:2012ra} 
  D.~Hernandez and A.~Y.~Smirnov,
  %``Lepton mixing and discrete symmetries,''
  Phys.\ Rev.\ D {\bf 86}, 053014 (2012)
  [arXiv:1204.0445 [hep-ph]].
  %%CITATION = ARXIV:1204.0445;%%

%\cite{Luhn:2013lkn}
\bibitem{Luhn:2013lkn} 
  C.~Luhn,
  %``Trimaximal TM$_{1}$ neutrino mixing in S$_{4}$ with spontaneous CP violation,''
  Nucl.\ Phys.\ B {\bf 875}, 80 (2013)
  [arXiv:1306.2358 [hep-ph]].
  %%CITATION = ARXIV:1306.2358;%%

\bibitem{Z2CP4}
%\cite{Hanlon:2013ska}
%\bibitem{Hanlon:2013ska} 
  A.~D.~Hanlon, S.~F.~Ge and W.~W.~Repko,
  %``Phenomenological consequences of residual $ \mathbb{Z}^s_2$ and $ \overline {\mathbb{Z}}^s_2$ symmetries,''
  Phys.\ Lett.\ B {\bf 729}, 185 (2014)
  [arXiv:1308.6522 [hep-ph]].
  %%CITATION = ARXIV:1308.6522;%%


%%%%%%%%%%%%%%%%%%%%% Group Scans

\bibitem{scans0}
%\cite{Parattu:2010cy}
%\bibitem{Parattu:2010cy} 
  K.~M.~Parattu and A.~Wingerter,
  %``Tribimaximal Mixing From Small Groups,''
  Phys.\ Rev.\ D {\bf 84}, 013011 (2011)
  [arXiv:1012.2842 [hep-ph]];
  %%CITATION = ARXIV:1012.2842;%%
%\cite{Lam:2012ga}
%\bibitem{Lam:2012ga} 
  C.~S.~Lam,
  %``Finite Symmetry of Leptonic Mass Matrices,''
  Phys.\ Rev.\ D {\bf 87}, no. 1, 013001 (2013)
  [arXiv:1208.5527 [hep-ph]];
  %%CITATION = ARXIV:1208.5527;%%
%\cite{Holthausen:2012wt}
%\bibitem{Holthausen:2012wt} 
  M.~Holthausen, K.~S.~Lim and M.~Lindner,
  %``Lepton Mixing Patterns from a Scan of Finite Discrete Groups,''
  Phys.\ Lett.\ B {\bf 721}, 61 (2013)
  [arXiv:1212.2411 [hep-ph]];
  %%CITATION = ARXIV:1212.2411;%%
%\cite{Fonseca:2014koa}
%\bibitem{Fonseca:2014koa} 
  R.~M.~Fonseca and W.~Grimus,
  %``Classification of lepton mixing matrices from finite residual symmetries,''
  JHEP {\bf 1409}, 033 (2014)
  [arXiv:1405.3678 [hep-ph]].
  %%CITATION = ARXIV:1405.3678;%%
%\cite{Talbert:2014bda}


\bibitem{scans1}
%\bibitem{Talbert:2014bda} 
  J.~Talbert,
  %``[Re]constructing Finite Flavour Groups: Horizontal Symmetry Scans from the Bottom-Up,''
  JHEP {\bf 1412}, 058 (2014)
  [arXiv:1409.7310 [hep-ph]].
  %%CITATION = ARXIV:1409.7310;%%


%%%%%%%%%%%%%%%%%%%%%%%Golden Ratio Mixing

\bibitem{GRmixing0}
%\bibitem{Datta:2003qg}
  A.~Datta, F.~S.~Ling and P.~Ramond,
  %``Correlated hierarchy, Dirac masses and large mixing angles,''
  Nucl.\ Phys.\ B {\bf 671}, 383 (2003)
  [hep-ph/0306002]. 


\bibitem{GRmixing1}
%\cite{Kajiyama:2007gx}
%\bibitem{Kajiyama:2007gx} 
  Y.~Kajiyama, M.~Raidal and A.~Strumia,
  %``The Golden ratio prediction for the solar neutrino mixing,''
  Phys.\ Rev.\ D {\bf 76}, 117301 (2007)
  [arXiv:0705.4559 [hep-ph]];
  %%CITATION = ARXIV:0705.4559;%%

\bibitem{A5GRmodels}
%\cite{Everett:2008et}
%\bibitem{Everett:2008et} 
  L.~L.~Everett and A.~J.~Stuart,
  %``Icosahedral (A(5)) Family Symmetry and the Golden Ratio Prediction for Solar Neutrino Mixing,''
  Phys.\ Rev.\ D {\bf 79}, 085005 (2009)
  [arXiv:0812.1057 [hep-ph]];
  %%CITATION = ARXIV:0812.1057;%%
%\cite{Feruglio:2011qq}
%\bibitem{Feruglio:2011qq} 
  F.~Feruglio and A.~Paris,
  %``The Golden Ratio Prediction for the Solar Angle from a Natural Model with A5 Flavour Symmetry,''
  JHEP {\bf 1103}, 101 (2011)
  [arXiv:1101.0393 [hep-ph]];
  %%CITATION = ARXIV:1101.0393;%%
%\cite{Ding:2011cm}
%\bibitem{Ding:2011cm} 
  G.~J.~Ding, L.~L.~Everett and A.~J.~Stuart,
  %``Golden Ratio Neutrino Mixing and $A_5$ Flavor Symmetry,''
  Nucl.\ Phys.\ B {\bf 857}, 219 (2012)
  [arXiv:1110.1688 [hep-ph]].
  %%CITATION = ARXIV:1110.1688;%%


\bibitem{kingluhnreview}
%\cite{King:2013eh}
%\bibitem{King:2013eh} 
  S.~F.~King and C.~Luhn,
  %``Neutrino Mass and Mixing with Discrete Symmetry,''
  Rept.\ Prog.\ Phys.\  {\bf 76}, 056201 (2013)
  [arXiv:1301.1340 [hep-ph]].
  %%CITATION = ARXIV:1301.1340;%%

%%%%%%%%%%%%Majorana phases
\bibitem{majorana1}
%\cite{Bilenky:1980cx}
%\bibitem{Bilenky:1980cx} 
  S.~M.~Bilenky, J.~Hosek and S.~T.~Petcov,
  %``On Oscillations of Neutrinos with Dirac and Majorana Masses,''
  Phys.\ Lett.\ B {\bf 94}, 495 (1980).
  %%CITATION = PHLTA,B94,495;%%
%\cite{Bilenky:1987ty}

\bibitem{majorana2}
%\cite{Schechter:1980gr}
%\bibitem{Schechter:1980gr} 
  J.~Schechter and J.~W.~F.~Valle,
  %``Neutrino Masses in SU(2) x U(1) Theories,''
  Phys.\ Rev.\ D {\bf 22}, 2227 (1980);
  %%CITATION = PHRVA,D22,2227;%%
%\cite{Doi:1980yb}
%\bibitem{Doi:1980yb} 
  M.~Doi, T.~Kotani, H.~Nishiura, K.~Okuda and E.~Takasugi,
  %``CP Violation in Majorana Neutrinos,''
  Phys.\ Lett.\ B {\bf 102}, 323 (1981).
  %%CITATION = PHLTA,B102,323;%%


%%%%%%%%%%%%%%%%%%%%%%%%%%%%%%%%%%%%

\bibitem{Jarlskog}
 %\cite{Jarlskog:1985ht}
%\bibitem{Jarlskog:1985ht} 
  C.~Jarlskog,
  %``Commutator of the Quark Mass Matrices in the Standard Electroweak Model and a Measure of Maximal CP Violation,''
  Phys.\ Rev.\ Lett.\  {\bf 55}, 1039 (1985).
  %%CITATION = PRLTA,55,1039;%%


%%%%%%%%%%%%%%%%%%Weak basis invariants
\bibitem{WBinv1}
%\cite{Branco:1986gr}
%\bibitem{Branco:1986gr} 
  G.~C.~Branco, L.~Lavoura and M.~N.~Rebelo,
  %``Majorana Neutrinos and {CP} Violation in the Leptonic Sector,''
  Phys.\ Lett.\ B {\bf 180}, 264 (1986);
  %%CITATION = PHLTA,B180,264;%%

\bibitem{WBinv2}
%\cite{Nieves:1987pp}
%\bibitem{Nieves:1987pp} 
  J.~F.~Nieves and P.~B.~Pal,
  %``Minimal Rephasing Invariant {CP} Violating Parameters With Dirac and Majorana Fermions,''
  Phys.\ Rev.\ D {\bf 36}, 315 (1987).
  %%CITATION = PHRVA,D36,315;%%

\bibitem{WBinv3}
%\cite{delAguila:1995bk}
%\bibitem{delAguila:1995bk} 
  F.~del Aguila and M.~Zralek,
  %``CP violation in the lepton sector with Majorana neutrinos,''
  Nucl.\ Phys.\ B {\bf 447}, 211 (1995)
  [hep-ph/9504228].
  %%CITATION = HEP-PH/9504228;%%

\bibitem{WBinv4}
%\cite{delAguila:1996ex}
%\bibitem{delAguila:1996ex} 
  F.~del Aguila, J.~A.~Aguilar-Saavedra and M.~Zralek,
  %``Invariant analysis of CP violation,''
  Comput.\ Phys.\ Commun.\  {\bf 100}, 231 (1997)
  [hep-ph/9607311].
  %%CITATION = HEP-PH/9607311;%%

\bibitem{WBinv5}
%\cite{Branco:1998bw}
%\bibitem{Branco:1998bw} 
  G.~C.~Branco, M.~N.~Rebelo and J.~I.~Silva-Marcos,
  %``Degenerate and quasidegenerate Majorana neutrinos,''
  Phys.\ Rev.\ Lett.\  {\bf 82}, 683 (1999)
  [hep-ph/9810328].
  %%CITATION = HEP-PH/9810328;%%

\bibitem{WBinv6}
%\cite{AguilarSaavedra:2000vr}
%\bibitem{AguilarSaavedra:2000vr} 
  J.~A.~Aguilar-Saavedra and G.~C.~Branco,
  %``Unitarity triangles and geometrical description of CP violation with Majorana neutrinos,''
  Phys.\ Rev.\ D {\bf 62}, 096009 (2000)
  [hep-ph/0007025].
  %%CITATION = HEP-PH/0007025;%%

\bibitem{WBinv7}
%\cite{Nieves:2001fc}
%\bibitem{Nieves:2001fc} 
  J.~F.~Nieves and P.~B.~Pal,
  %``Rephasing invariant CP violating parameters with Majorana neutrinos,''
  Phys.\ Rev.\ D {\bf 64}, 076005 (2001)
  [hep-ph/0105305].
  %%CITATION = HEP-PH/0105305;%%

\bibitem{WBinv8}
%\cite{Branco:2004hu}% Mentions mass
%\bibitem{Branco:2004hu} 
  G.~C.~Branco and M.~N.~Rebelo,
  %``Leptonic CP violation and neutrino mass models,''
  New J.\ Phys.\  {\bf 7}, 86 (2005)
  [hep-ph/0411196].
  %%CITATION = HEP-PH/0411196;%%

\bibitem{WBinv9}
%\cite{Farzan:2006vj} Invariants in terms of neutrino mass matrix elements
%\bibitem{Farzan:2006vj} 
  Y.~Farzan and A.~Y.~Smirnov,
  %``Leptonic CP violation: Zero, maximal or between the two extremes,''
  JHEP {\bf 0701}, 059 (2007)
  [hep-ph/0610337].
  %%CITATION = HEP-PH/0610337;%%

\bibitem{WBinv10}
%\cite{Dreiner:2007yz}
%\bibitem{Dreiner:2007yz} 
  H.~K.~Dreiner, J.~S.~Kim, O.~Lebedev and M.~Thormeier,
  %``Supersymmetric Jarlskog invariants: The Neutrino sector,''
  Phys.\ Rev.\ D {\bf 76}, 015006 (2007)
  [hep-ph/0703074 [HEP-PH]].
  %%CITATION = HEP-PH/0703074;%%

\bibitem{WBinv11}
%\cite{Jenkins:2007ip} I1 and I2 defined in here
%\bibitem{Jenkins:2007ip} 
  E.~E.~Jenkins and A.~V.~Manohar,
  %``Rephasing Invariants of Quark and Lepton Mixing Matrices,''
  Nucl.\ Phys.\ B {\bf 792}, 187 (2008)
  [arXiv:0706.4313 [hep-ph]].
  %%CITATION = ARXIV:0706.4313;%%

\bibitem{WBinv12}
%\cite{Jenkins:2009dy}
%\bibitem{Jenkins:2009dy} 
  E.~E.~Jenkins and A.~V.~Manohar,
  %``Algebraic Structure of Lepton and Quark Flavor Invariants and CP Violation,''
  JHEP {\bf 0910}, 094 (2009)
  [arXiv:0907.4763 [hep-ph]].
  %%CITATION = ARXIV:0907.4763;%%

\bibitem{WBinv13}
%\cite{Branco:2011zb}Mentions masses
%\bibitem{Branco:2011zb} 
  G.~C.~Branco, R.~G.~Felipe and F.~R.~Joaquim,
  %``Leptonic CP Violation,''
  Rev.\ Mod.\ Phys.\  {\bf 84}, 515 (2012)
  [arXiv:1111.5332 [hep-ph]].
  %%CITATION = ARXIV:1111.5332;%%

%%%%%%%%%%%%%%%%%%%%%%%%END OF WEAK BASIS INV. REFERENCES


\bibitem{Dingnew}
%\cite{Chen:2014wxa}
%\bibitem{Chen:2014wxa} 
  P.~Chen, C.~C.~Li and G.~J.~Ding,
  %``Lepton Flavor Mixing and CP Symmetry,''
  arXiv:1412.8352 [hep-ph].
  %%CITATION = ARXIV:1412.8352;%%



%%%%%%%%%%%%%%%%%%%% RGE
\bibitem{CLRGECN}
  %%CITATION = ARXIV:0712.3759;%%
%\cite{Antusch:2007ib}
%\bibitem{Antusch:2007ib} 
  S.~Antusch, S.~F.~King and M.~Malinsky,
  %``Third Family Corrections to Tri-bimaximal Lepton Mixing and a New Sum Rule,''
  Phys.\ Lett.\ B {\bf 671}, 263 (2009)
  [arXiv:0711.4727 [hep-ph]];
  %%CITATION = ARXIV:0711.4727;%%
%\cite{Antusch:2007vw}
%\bibitem{Antusch:2007vw} 
  S.~Antusch, S.~F.~King and M.~Malinsky,
  %``Third Family Corrections to Quark and Lepton Mixing in SUSY Models with non-Abelian Family Symmetry,''
  JHEP {\bf 0805}, 066 (2008)
  [arXiv:0712.3759 [hep-ph]];
%\cite{Boudjemaa:2008jf}
%\bibitem{Boudjemaa:2008jf} 
  S.~Boudjemaa and S.~F.~King,
  %``Deviations from Tri-bimaximal Mixing: Charged Lepton Corrections and Renormalization Group Running,''
  Phys.\ Rev.\ D {\bf 79}, 033001 (2009)
  [arXiv:0808.2782 [hep-ph]];
  %%CITATION = ARXIV:0808.2782;%%
%\cite{Antusch:2008yc}
%\bibitem{Antusch:2008yc} 
  S.~Antusch, S.~F.~King and M.~Malinsky,
  %``Perturbative Estimates of Lepton Mixing Angles in Unified Models,''
  Nucl.\ Phys.\ B {\bf 820}, 32 (2009)
  [arXiv:0810.3863 [hep-ph]].
  %%CITATION = ARXIV:0810.3863;%%

%%%%%%%%%%%%%%%%%%%%%General RGE
\bibitem{genrge}
%\cite{Casas:1999tg}
%\bibitem{Casas:1999tg} 
  J.~A.~Casas, J.~R.~Espinosa, A.~Ibarra and I.~Navarro,
  %``General RG equations for physical neutrino parameters and their phenomenological implications,''
  Nucl.\ Phys.\ B {\bf 573}, 652 (2000)
  [hep-ph/9910420];
  %%CITATION = HEP-PH/9910420;%%
%\cite{Chankowski:2001mx}
%\bibitem{Chankowski:2001mx} 
  P.~H.~Chankowski and S.~Pokorski,
  %``Quantum corrections to neutrino masses and mixing angles,''
  Int.\ J.\ Mod.\ Phys.\ A {\bf 17}, 575 (2002)
  [hep-ph/0110249];
  %%CITATION = HEP-PH/0110249;%%
%\cite{Antusch:2003kp}
%\bibitem{Antusch:2003kp} 
  S.~Antusch, J.~Kersten, M.~Lindner and M.~Ratz,
  %``Running neutrino masses, mixings and CP phases: Analytical results and phenomenological consequences,''
  Nucl.\ Phys.\ B {\bf 674}, 401 (2003)
  [hep-ph/0305273].
  %%CITATION = HEP-PH/0305273;%%

%%%%%%%%%%%%%%%%%%%% CN only
\bibitem{CN}
%\cite{King:2003xq}
%\bibitem{King:2003xq} 
  S.~F.~King and I.~N.~R.~Peddie,
  %``Canonical normalization and Yukawa matrices,''
  Phys.\ Lett.\ B {\bf 586}, 83 (2004)
  [hep-ph/0312237].
  %%CITATION = HEP-PH/0312237;%%

%%%%%%%%%%%%%%%%%%%%% CL
\bibitem{CL}
%\cite{Petcov:2014laa}
%\bibitem{Petcov:2014laa} 
  S.~T.~Petcov,
  %``Predicting the Values of the Leptonic CP Violation Phases,''
  arXiv:1405.6006 [hep-ph];
%\cite{Ballett:2014dua}
%\bibitem{Ballett:2014dua} 
  P.~Ballett, S.~F.~King, C.~Luhn, S.~Pascoli and M.~A.~Schmidt,
  %``Testing solar lepton mixing sum rules in neutrino oscillation experiments,''
  arXiv:1410.7573 [hep-ph];
%\cite{Girardi:2014faa}
%\bibitem{Girardi:2014faa} 
  I.~Girardi, S.~T.~Petcov and A.~V.~Titov,
  %``Determining the Dirac CP Violation Phase in the Neutrino Mixing Matrix from Sum Rules,''
  arXiv:1410.8056 [hep-ph].
  %%CITATION = ARXIV:1410.8056;%%

\bibitem{TBmix} 
%\cite{Harrison:2002er}
%\bibitem{Harrison:2002er} 
  P.~F.~Harrison, D.~H.~Perkins and W.~G.~Scott,
  %``Tri-bimaximal mixing and the neutrino oscillation data,''
  Phys.\ Lett.\ B {\bf 530}, 167 (2002)
  [hep-ph/0202074];
  %%CITATION = HEP-PH/0202074;%%
%\cite{Harrison:2002kp}
%\bibitem{Harrison:2002kp} 
  P.~F.~Harrison and W.~G.~Scott,
  %``Symmetries and generalizations of tri - bimaximal neutrino mixing,''
  Phys.\ Lett.\ B {\bf 535}, 163 (2002)
  [hep-ph/0203209];
  %%CITATION = HEP-PH/0203209;%%
%\cite{Xing:2002sw}
%\bibitem{Xing:2002sw} 
  Z.~z.~Xing,
  %``Nearly tri bimaximal neutrino mixing and CP violation,''
  Phys.\ Lett.\ B {\bf 533}, 85 (2002)
  [hep-ph/0204049];
  %%CITATION = HEP-PH/0204049;%%
%\cite{He:2003rm}
%\bibitem{He:2003rm} 
  X.~G.~He and A.~Zee,
  %``Some simple mixing and mass matrices for neutrinos,''
  Phys.\ Lett.\ B {\bf 560}, 87 (2003)
  [hep-ph/0301092].
  %%CITATION = HEP-PH/0301092;%%







%%%%%%%%%%%%%Delta(96) and BT mixing
\bibitem{BTmixing}
%\cite{Toorop:2011jn}
%\bibitem{Toorop:2011jn} 
  R.~d.~A.~Toorop, F.~Feruglio and C.~Hagedorn,
  %``Discrete Flavour Symmetries in Light of T2K,''
  Phys.\ Lett.\ B {\bf 703}, 447 (2011)
  [arXiv:1107.3486 [hep-ph]];
  %%CITATION = ARXIV:1107.3486;%%
%\cite{Ding:2012xx}
%\bibitem{Ding:2012xx} 
  G.~J.~Ding,
  %``TFH Mixing Patterns, Large $\theta_{13}$ and $\Delta(96)$ Flavor Symmetry,''
  Nucl.\ Phys.\ B {\bf 862}, 1 (2012)
  [arXiv:1201.3279 [hep-ph]];
  %%CITATION = ARXIV:1201.3279;%%
%\cite{King:2012in}
%\bibitem{King:2012in} 
  S.~F.~King, C.~Luhn and A.~J.~Stuart,
  %``A Grand Delta(96) x SU(5) Flavour Model,''
  Nucl.\ Phys.\ B {\bf 867}, 203 (2013)
  [arXiv:1207.5741 [hep-ph]].
  %%CITATION = ARXIV:1207.5741;%%









\end{thebibliography}
\end{document}